\documentclass[11pt,a4paper]{article}

\pdfoutput=1
\usepackage{jcappub} 
\usepackage[T1]{fontenc} 
\usepackage[utf8]{inputenc}
\usepackage{pstool}
\usepackage{hyperref}
\usepackage{amsmath,amssymb}
\usepackage{float}
\usepackage{graphicx}
\usepackage{graphics}
\usepackage{bm}
\usepackage{latexsym}
\usepackage{psfrag}
\usepackage{color}
\usepackage{subfigure}
\usepackage[normalem]{ulem}
\usepackage{textcomp}
\usepackage{comment}
\usepackage{mathalpha}
\usepackage{tabularx}
\usepackage{placeins}   

\makeatletter
\gdef\@fpheader{}
\makeatother

\def\bea{\begin{eqnarray}}
\def\eea{\end{eqnarray}}

\def\f{\frac}

\def\d{{\rm d}}
\def\Mpl{M_{_{\mathrm{Pl}}}}
\def\mpcinv{{\rm Mpc}^{-1}}

\def\ps{\mathcal{P}_{_{\mathrm{S}}}}
\def\pt{\mathcal{P}_{_{\mathrm{T}}}}
\def\barpow1d{\mathcal{P}^{1\rm D}_{\mathrm{b}}}
\def\pmrd{\Delta^{1\rm D}_{\mathrm{m}}}
\def\ns{n_{\mathrm{s}}}

\def\fnl{f_{_{\rm NL}}}

\newcommand{\HI}{\rm HI}

\newcommand{\K}{\rm K}

\allowdisplaybreaks

\begin{document}

\title{Constraining ultra slow roll inflation using cosmological datasets}

\author{H. V. Ragavendra$^{1\,\ast}$,}
\emailAdd{$^\ast$ragavendra@rrimail.rri.res.in}
\affiliation{$^1$Raman Research Institute, C.~V.~Raman Avenue, Sadashivanagar, 
Bengaluru~560080, India}
\author{Anjan Kumar Sarkar$^{2\,\dagger}$ and}
\emailAdd{$^\dagger$asarkar@ncra.tifr.res.in}
\affiliation{$^2$National Centre for Radio Astrophysics, TIFR, Pune University 
Campus, Post Bag 3, Pune 411 007, India.}
\author{Shiv K. Sethi$^{1\,\ddagger}$}
\emailAdd{$^\ddagger$sethi@rri.res.in}

\abstract
{In recent years, the detection of gravitational waves by LIGO and PTA collaborations 
have raised the intriguing possibility of excess matter power at small scales. Such an 
increase can be achieved by ultra slow roll (USR) phase during inflationary epoch. 
We constrain excess power over small scales within the framework of such models using 
cosmological datasets, particularly of CMB anisotropies and Lyman-$\alpha$. 
We parameterize the USR phase in terms of the 
e-fold at the onset of USR (counted from the end of inflation) $\bar N_1$ and 
the duration of USR phase $\Delta N$.
The former dictates the scale of enhancement in the primordial power spectrum, while the latter determines the amplitude of such an
enhancement.
From a joint dataset of CMB and galaxy surveys, we obtain 
$\bar N_1 \lesssim 45$ with no bound on $\Delta N$. This in turn implies that 
the scales over which the power spectrum can deviate significantly from the 
nearly scale invariant behavior of a typical slow-roll model is 
$k \gtrsim 1 \, \rm Mpc^{-1}$.
On the other hand, the Lyman-$\alpha$ data is sensitive to baryonic power 
spectrum along the line of sight. We consider a semi-analytic theoretical method 
and high spectral-resolution Lyman-$\alpha$ data to constrain the model. 
The Lyman-$\alpha$ data limits both the USR parameters: 
$\bar N_1 \lesssim 41$ and $\Delta N \lesssim 0.4$. 
This constrains the amplitude of the power spectrum enhancement to be less than 
a factor of hundred over scales $1 \lesssim k/\mpcinv \lesssim 100$, 
thereby considerably improving the constraint on power over these scales as 
compared to the bounds arrived at from CMB  spectral distortion.}

\maketitle

\section{Introduction}

The current cosmological data bears out the $\Lambda$CDM model at large scales ($k \lesssim 0.2 \, \rm Mpc^{-1}$). At smaller scales, the situation remains unsettled. 
While a set of observables suggest decrement of matter power at small scales, the 
enhancement of primordial scalar power at small scales is also a feature that has
been sought after in the recent literature. Such enhancement leads to production
of primordial black holes (PBHs) and secondary gravitational waves (GWs) and so 
models of primordial universe that effect such a feature have been constructed and 
examined extensively (see for reviews, Refs.~\cite{Ozsoy:2023ryl,Domenech:2021ztg,LISACosmologyWorkingGroup:2023njw}).
This effort has been spurred by the detection of mergers of supermassive 
black-holes by {\tt LIGO} and, recently, with the data release of {\tt PTA} which 
claims the detection of stochastic GWs~\cite{LIGOScientific:2018mvr,LIGOScientific:2021usb,LIGOScientific:2021djp,NANOGrav:2023gor,EPTA:2023fyk}. 
It has been suggested that the black-holes of the {\tt LIGO} merger events 
are primordial black holes (see Refs.~\cite{Bird:2016dcv,Sasaki:2016jop,DeLuca:2020qqa,Jedamzik:2020ypm,Jedamzik:2020omx,Chen:2024dxh} and references therein).
The data of {\tt PTA} suggests that the scalar-induced secondary GWs seem to 
be the best-fit candidate to explain the signal~\cite{NANOGrav:2023hvm,
Cai:2023dls,Vagnozzi:2023lwo,Choudhury:2023kam,Firouzjahi:2023lzg,Das:2023nmm,Liu:2023pau,Ellis:2023oxs,Gangopadhyay:2023qjr,Fei:2023iel,Choudhury:2024one,Chen:2024twp}.

Amongst the many models of early universe that may lead to the desired enhancement
of scalar power over small scales, inflationary models with a brief epoch of 
ultra slow roll (USR) seem to be the ones that effectively achieve the required 
feature. Such models are realised by a variety of potentials with features such as 
(near-)inflection points, a break, a bump or a dip in them~\cite{Starobinsky:1992ts,Hazra:2010ve,Martin:2014kja,Garcia-Bellido:2017mdw,Germani:2017bcs,Ballesteros:2017fsr,Ezquiaga:2017fvi,Bezrukov:2017dyv,Drees:2019xpp,Atal:2019cdz,Mishra:2019pzq,Ballesteros:2020qam,Kefala:2020xsx,Braglia:2020eai,ZhengRuiFeng:2021zoz,Braglia:2022phb,Zhao:2023zbg,Choudhury:2023hfm}~(see 
also~\cite{Ragavendra:2023ret} for a brief review of such potentials). 
There are also attempts at reconstructing this model using the characteristic behavior 
of the first slow roll parameter~\cite{Byrnes:2018txb,Motohashi:2019rhu,Tasinato:2020vdk,Ragavendra:2020sop,Franciolini:2022pav,Domenech:2023dxx}.
USR models while amplifying the scalar power, also lead to certain tell-tale features 
in the spectrum. Their scalar power spectrum contains a unique dip in the amplitude
just prior to the sharp rise in power~\cite{Ozsoy:2019lyy,Balaji:2022zur}\footnote{
This dip is at the tree-level power spectrum and may be modified or filled in by
loop-level contributions~\cite{Franciolini:2023lgy}.}. 
The rise is 
typically of shape $k^4$ until the peak amplitude is achieved~\cite{Byrnes:2018txb,Ng:2021hll,Cole:2022xqc}.
The shape post the peak is determined by the dynamics of inflation after the phase of USR.

The crucial aspects in such a spectrum are the location of the peak and the amount 
of enhancement of power over the typical nearly scale-invariant amplitude over large
scales. These features are determined by the onset and duration of the USR phase 
during inflation.
In the context of production of PBHs and secondary GWs, the location of peak dictates 
the mass of PBHs and the frequency of the peak amplitude in the spectral density of
GWs, denoted as $\Omega_{\rm GW}$. The amount of enhancement in turn determines the 
population of PBHs and the amplitude of $\Omega_{\rm GW}$ at its maximum.
Thus the onset and duration of USR have a direct impact on the characteristics of 
the relevant observables.

Since the onset of USR dictates the relevant mass of PBHs or frequency of GWs,
it is usually tuned to values as per the desired range. The duration of USR is also 
tuned to achieve the required population fraction of PBHs or the amplitude of 
GWs. The production of PBHs is exponentially sensitive to the peak amplitude of 
scalar power, and therefore on the duration of USR. Hence, these parameters are 
fine-tuned in models that attempt to achieve significant fraction of PBHs along 
with secondary GWs.

There have been attempts in the literature to examine the consistency of models 
leading to enhanced power over small scales and constraints from CMB anisotropies 
over large scales. In Refs.~\cite{Byrnes:2018txb,Bhaumik:2019tvl,Ragavendra:2020sop,Iacconi:2021ltm,Cicoli:2022sih,Karam:2022nym,Balaji:2022zur,Qin:2023lgo,Tagliazucchi:2023dai}, 
there are discussions about the value of scalar spectral index and tensor-to-scalar 
ratio arising from such models over large scales and how they are often in tension 
with the CMB data.
It is also noted that there can be strong running of spectral index when the enhancement
in power is achieved closer to CMB scales~\cite{Kohri:2007qn,Balaji:2022zur}.
Ref.~\cite{Qin:2023lgo} specifically focuses on a multi-field model of inflation 
and, using an analytical template of power spectrum, arrives at bounds on parameters of 
the two-field potential through a Markov-chain Monte-Carlo (MCMC) analysis against CMB 
dataset. However, there are no rigorous constraints on the models of USR through similar 
analysis. 
In this work, we attempt to constrain the dynamics of USR, particularly its onset and 
duration during inflation, using MCMC analysis against a variety of datasets over large 
scales.

Before the details of our analysis, we briefly summarize some of the key outcomes 
of the  USR phase, based on the behavior of various inflationary potentials leading to USR. 
Over the course of about $70$ e-folds of inflation, we need about four 
decades of CMB scales to have a nearly scale-invariant spectrum with reasonably
small tensor-to-scalar ratio for agreement with Planck data. Hence we may expect
just the initial $14\hbox{--}16$ e-folds 
to have slow-roll dynamics during which these scales can evolve from sufficiently 
deep inside the Hubble radius and grow into the super-Hubble regime. 
This allows USR to occur anywhere in the last $55$ e-folds of inflation.
This essentially leaves a wide window of e-folds where USR can occur and last.
On the other hand, a broad bound on the duration of USR comes from the limit on 
the enhancement of power spectrum $\ps(k) < 1$\,. Since $\ps(k) \propto \exp(6 \Delta N)$ 
($\Delta N$ denoting the duration in e-folds) during USR, this restricts the duration 
to be less than about three e-folds. 
Besides, if USR were to occur over an epoch that enhances power around scales of
$k \sim 10^4\,\mpcinv$\,, then the bound on $\ps(k) < 10^{-5}$ from spectral distortions
restricts USR to be less than $1.5$ e-folds in this particular regime~\cite{Mather:1993ij,Fixsen:1996nj}.
Beyond these broad bounds, there are no strong constraints on the epoch of USR.

In this work, we attempt to directly constrain  the epoch of the earliest onset of USR and
its duration during inflation. To achieve this, we characterize the model of USR using a 
general parameterization of the first slow roll parameter over the entire course of inflation. 
This parameterization captures the essential features of USR models at the level of 
background and allows us to compute the dynamics of first-order perturbations.
Moreover, it has explicit parameters that determine the onset of USR phase, 
its duration, and the smoothness of transition from SR to USR phase. 
We discuss the setup and details of this model in section~\ref{sec:USR}.
In section~\ref{sec:res}, we test our models against data on a range of scales, in 
particular the CMB and Lyman-$\alpha$ datasets.
We first perform a comparison of model against the data of CMB anisotropies from 
{\tt Planck} and {\tt BICEP/KECK} missions.
We then include the datasets of galaxy surveys from {\tt BOSS} and {\tt DES} missions, 
along with CMB to inspect the improvement in the constraints. We observe that CMB
anisotropy data provide the essential bounds on the parameters and they are mildly
improved by the later additions. 
We discuss this part of data analysis in section~\ref{sec:cmbplus}.
To test our theoretical  model against small-scale data, we use the evolution of effective optical depth from high spectral-resolution Lyman-$\alpha$ data \cite{faucher2008direct}. The details of this method and the results of statistical comparison are given in section~\ref{sec:lyalpha}. Lastly, we discuss the crucial takeaways and outlook of 
our work in section~\ref{sec:conc}.


\section{Model of ultra-slow roll inflation}\label{sec:USR}

There are several potentials in the literature that achieve a brief epoch of USR
during inflation (for a review of USR, refer~\cite{Ragavendra:2023ret}). 
These models have a phase where the canonical scalar field (inflaton $\phi$) 
undergoes a phase of exponential decrease in the velocity in terms of e-folds $N$. 
Such dynamics of inflaton field is best studied through the behavior of the first 
slow roll parameter $\epsilon_1$. 
This parameter is defined with respect to the Hubble parameter $H$ as 
\begin{equation}
\epsilon_1(N) = -\frac{1}{H^2}\frac{\d H}{\d t}\,,
\label{eq:eps1-HN}
\end{equation}
where $t$ is the cosmic time. 
Using Friedmann equations in this case of inflation driven by a single canonical 
scalar field, it can be expressed as
\begin{equation}
\epsilon_1(N) = \frac{1}{2H^2}\left(\frac{\d \phi}{\d t}\right)^2
= \frac{1}{2}\left(\frac{\d \phi}{\d N}\right)^2\,.
\label{eq:eps1-phin}
\end{equation}
Hence the phase of USR is defined by the exponential decrease of $\epsilon_1$ in terms
of e-folds $N$, contrary to the SR phase where it evolves much slowly.

A parameterization of $\epsilon_1(N)$ that captures this behavior suitably is 
convenient to understand the dynamics and also the corresponding effect on perturbations.
There have been attempts at reconstruction of a brief epoch of USR during 
inflation, through parameterization of $\epsilon_1(N)$ in the 
literature~\cite{Tasinato:2020vdk,Motohashi:2019rhu,Franciolini:2022pav}. 
Such parametric models of $\epsilon_1(N)$ are sought to generalize the dynamics of USR,
without restricting to any specific potentials. We work with one such parametric 
model where the form of $\epsilon_1(N)$ is given by~\cite{Ragavendra:2020sop,Ragavendra:2023ret}
\begin{eqnarray}
\epsilon_1(N) &=& \f{\epsilon_{1_{\rm i}}\,e^{\epsilon_{2_{\rm i}}N}}{2}\,
\left[1 - \tanh\left(\f{N - N_1}{\delta N_1}\right)\right]
+ \f{\epsilon_{1_{\rm f}}}{2}\,
\left[1 + \tanh\left(\f{N - N_1}{\delta N_1}\right)\right]
+ e^{\f{N - N_{\rm e}}{\delta N_{\rm e}}}\,.
\label{eq:eps1-parametric}
\end{eqnarray}
In this parametrization, the onset of USR is determined by $N_1$ and the
smoothness of transition between USR and SR phases is determined by $\delta N_1$.
The duration of USR phase is controlled by $\Delta N$ which enters the model through
$\epsilon_{1_{\rm f}}$, as we set
$\epsilon_{1_{\rm f}}=\epsilon_{1_{\rm i}}\exp(-6\,\Delta N)$.
The onset and duration of USR, $N_1$ and $\Delta N$, are the two essential parameters
that we shall be varying to observe the effects on the scalar power over small scales.
Thus our parametric model gives us a direct handle on the USR dynamics
through the parameters $N_1$ and $\Delta N$\,.
We set the value of smoothness to be $\delta N_1 = 0.31$\,.
This choice of value corresponds to setting $\epsilon_2 \simeq -6$. 
This is the typical value achieved by $\epsilon_2$ in USR models driven by various 
potentials (refer~\cite{Ragavendra:2023ret} for details).
Variation of this parameter alters the value of $\epsilon_2$ and in turn affects
the steepness of rise in scalar power. We retain this parameter at its fixed value 
throughout our analysis.

The transition to USR phase is effected at $N_1$ by the terms containing hyperbolic 
tangents. The value of $\epsilon_1$ drops from $\epsilon_{1_{\rm i}}$ to 
$\epsilon_{1_{\rm f}}$ during the USR phase. 
Once it settles to $\epsilon_{1_{\rm f}}$, the final phase of SR begins and it 
lasts until the last term starts playing a role.
The end of inflation is achieved by making $\epsilon_1$ reach unity using the 
last term. We achieve $72$ e-folds of inflation by setting $N_{\rm e} = 72$ and 
the rapidity of the rise at the end is fixed as $\delta N_{\rm e} = 0.55$.

The evolution of other background quantities such as the inflaton $\phi(N)$ and
Hubble parameter $H(N)$ can be obtained from the behavior of $\epsilon_1(N)$ 
by inverting the relations~\eqref{eq:eps1-HN} and~\eqref{eq:eps1-phin} as 
\bea
\phi(N) &=& \phi_{\rm i} - \int^N_0 \d N'\,\sqrt{2\epsilon_1(N')}\,\Mpl\,, \\
H(N) &=& H_{\rm i}\exp\left[-\int^N_0 \d N' \epsilon_1(N') \right]\,,
\label{eq:H-evol}
\eea
along with suitable choice of initial conditions for respective quantities, namely
the field value $\phi_{\rm i}$ and the Hubble parameter $H_{\rm i}$.
The shape of the potential $V(\phi)$ can also be obtained from the behavior 
of $\phi(N)$ and $V(N) = [3-\epsilon_1(N)]\,H^2(N)\Mpl^2$\footnote{$\Mpl=\sqrt{\hbar c/(8\pi G)}$ is the reduced Planck mass in terms of the fundamental constants, the reduced Planck's constant $\hbar$, the speed of light $c$ and the gravitational constant $G$.}\,.

We numerically evolve the scalar and tensor perturbations over the background that 
is modelled through $\epsilon_1(N)$ and compute the corresponding spectra close 
to the end of inflation, well past the epoch of USR. 
The dimensionless power spectra of the scalar and tensor perturbations are defined,
in terms of the associated mode functions $f_k$ and $g_k$ respectively, as~\cite{Mukhanov:1990me,Martin:2004um,Kinney:2009vz,Baumann:2009ds,Sriramkumar:2009kg}
\begin{eqnarray}
\ps(k) &=& \f{k^3}{2\pi^2}\vert f_k \vert^2\,, \\
\pt(k) &=& 4\f{k^3}{2\pi^2}\vert g_k \vert^2\,.
\end{eqnarray}
The pivot scale $k_{\rm p}=0.05\,\mpcinv$ is set to exit the Hubble radius $50$ 
e-folds before the end of inflation, i.e., at $N_{\rm p}=22$ in our 
parametrization. Thus the scales of CMB $k=[10^{-4},1]\,\mpcinv$ are set to exit 
the Hubble radius during $N=[16,25]$ of our model.
The choices of $N_{\rm p}$, and $N_{\rm e}=72$ earlier, are made so as to let the largest 
observable scale (with $k \simeq 10^{-4}\,\mpcinv$) evolve from sufficiently deep inside the 
Hubble radius, for about $15$ e-folds. This duration also dilutes imprints of any preceding 
non-inflationary epoch that could affect the evolution of perturbations.
To relate to the convention of counting e-folds from the end of inflation, let us 
define 
\begin{equation}
\bar N = N_{\rm e} - N,
\end{equation}
which in our case becomes $\bar N = 72-N$\,. Thus, in terms of $\bar N$, the pivot scale
exits Hubble radius at $\bar N=50$ and the range corresponding to CMB scales is
$\bar N=[56,47]$\,. We shall use $\bar N$  in later analysis, when we interpret various 
bounds on e-folds.

\begin{figure}[!t]
\centering
\includegraphics[scale=0.30]{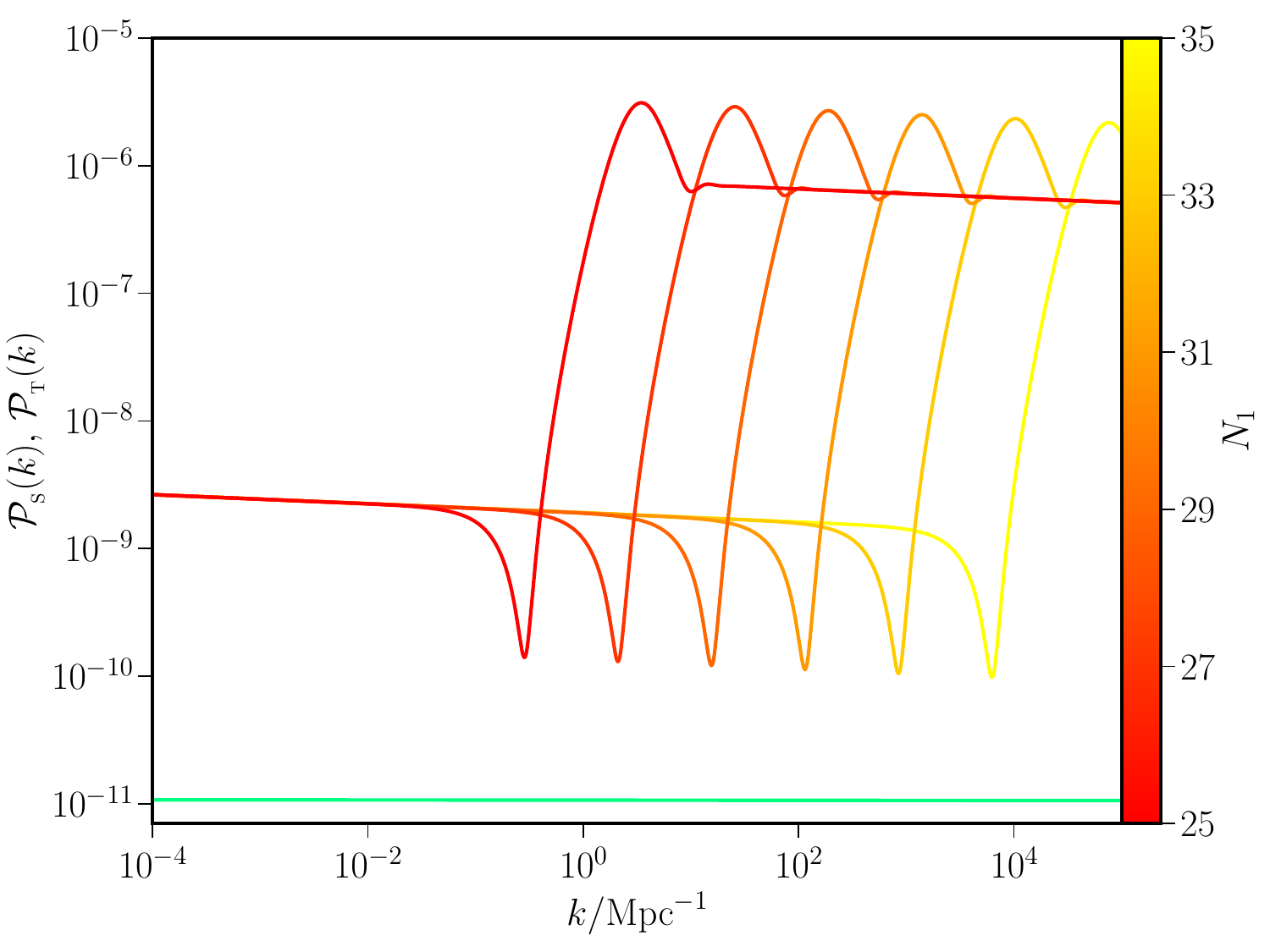}
\includegraphics[scale=0.30]{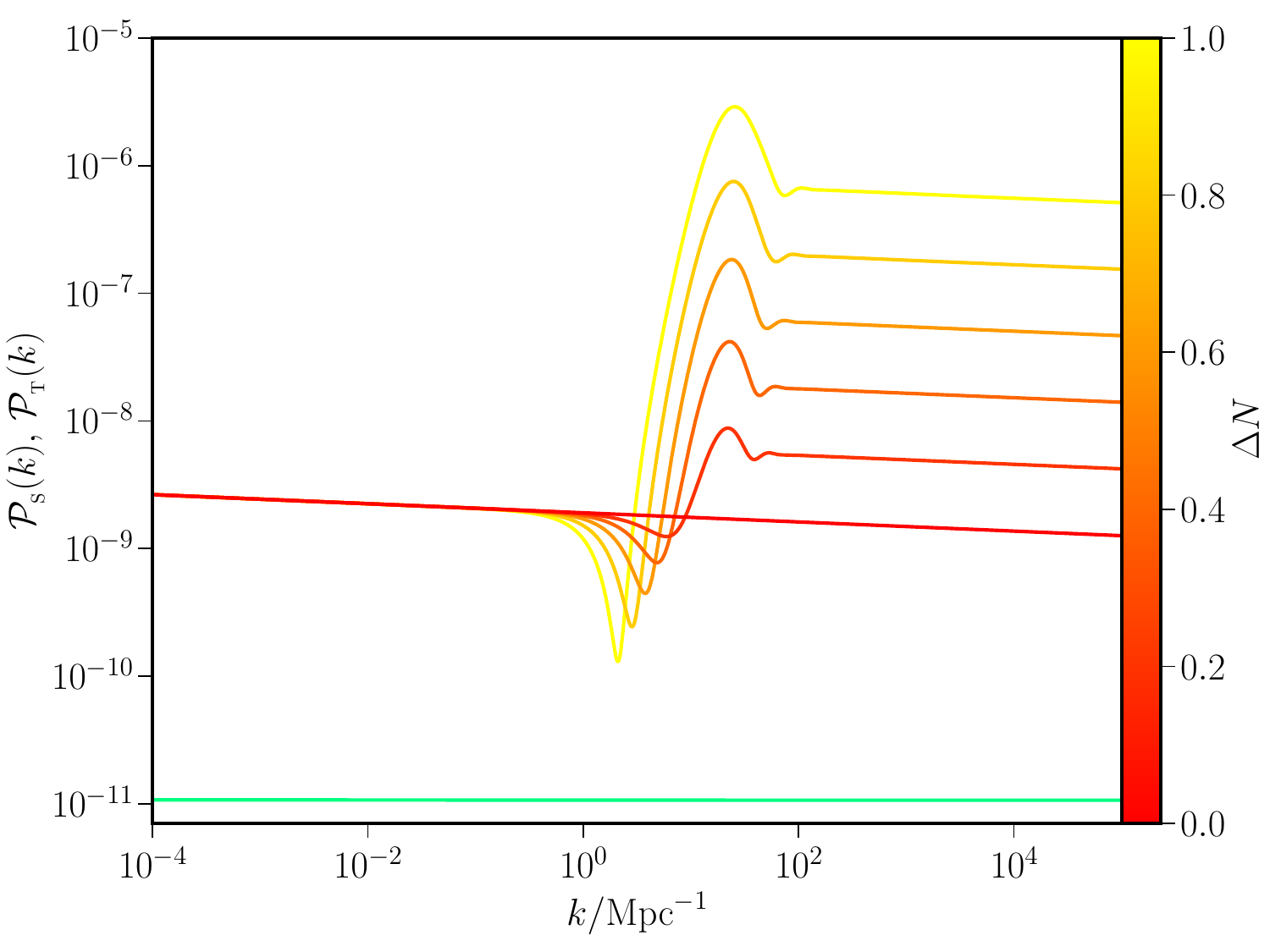}
\vskip -0.1in
\caption{We present the behavior of scalar power spectra (in shades of red to 
yellow) arising from the parametric model of USR with variation of the two 
parameters of interest: 
the onset of USR $N_1$ (in the left panel, with constant $\Delta N = 1$) and 
the duration of USR $\Delta N$ (in the right panel, with constant $N_1 = 27$). 
The location of enhancement of scalar power is determined by $N_1$ with larger 
values implying enhancement over smaller scales. The amplitude of enhancement 
is determined by $\Delta N$ with large values implying larger enhancement.
The range of parameters are chosen such that we preserve the nearly scale 
invariant behavior over $k \leq 0.1\,\mpcinv$ and also ensuring that we do not 
violate the upper bound on enhancement due to CMB spectral distortions over 
scales around $k \simeq 10^4\,\mpcinv$.
The tensor power spectrum remains unaffected by the USR dynamics and stays the
same across variation of parameters. It is displayed as a single horizontal 
green line.}
\label{fig:Ps-USR}
\end{figure}

The behavior of scalar power spectrum for the range of parameters we work with is 
presented in Fig.~\ref{fig:Ps-USR}. 
We observe that our model of $\epsilon_1(N)$ captures the essential features of the 
spectra arising from a generic model of USR achieved using a potential.
The scalar power spectrum contains the signatures of USR such as the dip, followed
by the rise of $k^4$, and a peak amplitude. Since our model has an extended duration
of SR after USR, the spectrum settles to another scale invariant regime past the peak. 
These individual features have been studied in several earlier works~\cite{Carrilho:2019oqg,Ozsoy:2018flq,Ozsoy:2019lyy,Karam:2022nym,Ozsoy:2021pws,Tasinato:2020vdk,Byrnes:2018txb,Motohashi:2019rhu,Liu:2020oqe,Ng:2021hll,Balaji:2022zur,Cole:2022xqc,Domenech:2023dxx}.

As mentioned earlier, since we are interested in enhancing scalar power over small scales,
the first phase of SR corresponds to the time when CMB scales exit the Hubble radius. 
The dynamics of this phase is determined by the prefactor in the first term of 
Eq.~\eqref{eq:eps1-parametric}, involving $\epsilon_{1_{\rm i}}$ and $\epsilon_{2_{\rm i}}$\,.
It is modelled to ensure proper values for the overall amplitude of the spectrum 
$A_{_{\rm S}}$, scalar spectral index $\ns$ and tensor-to-scalar ratio $r$ over 
scales exiting in the initial regime of SR.
Further, we also restrict the range of variation of $\Delta N$ so that the 
enhancement does not violate the bound on $\ps(k)$ due to spectral distortions in
CMB around $k \simeq 10^4\,\mpcinv$.
As we vary the parameter $N_1$ we observe that feature of dip and rise shift
across the range of wavenumber $k$, with larger values of $N_1$ shifting it to 
larger values of $k$\,. The increase in $\Delta N$ leads to increase in the
amplitude of enhancement while also making the dip more prominent prior to the
rise.

The tensor power spectrum remains nearly scale invariant throughout the range
of scales, as it is largely unaffected by USR dynamics.
However, they may suffer a step like drop in amplitude in case of models that 
have a brief rise in $\epsilon_1(N)$ just prior to the epoch of USR. Such models 
may even lead to an interruption in inflation if the rise in $\epsilon_1(N)$ 
crosses unity before decreasing rapidly during USR. These models are called punctuated
inflationary models~\cite{Jain:2008dw,Jain:2009pm,Qureshi:2016pjy,Ragavendra:2020old}. 
We do not include such a punctuation in our parametric model as we are mainly 
interested in the rise in the scalar power, though it can be included if required 
through an additional term in the parameterization (see Ref.~\cite{Ragavendra:2020sop} 
for such a reconstruction).

Based on our analysis, we obtain the following relation between the number of e-folds 
$N_1$ and the scale corresponding to the peak of the primordial power spectrum, 
$k_{\rm peak}$ 
(Figure~\ref{fig:Ps-USR}). 
For $\Delta N = 0.2$,
\begin{equation}
N_1 \simeq 25 + \ln\left(\f{k_{\rm peak}}{3\,\mpcinv}\right)
\label{eq:n1krel}
\end{equation}
and 
\begin{equation}
N_1 \simeq 25 + \ln\left(\f{k_{\rm peak}}{3.48\,\mpcinv}\right)
\label{eq:n1krel1}
\end{equation}
for $\Delta N = 1$.
From Fig.~\ref{fig:Ps-USR} it is clear that the power spectrum deviates 
significantly from the nearly scale invariant behavior over
$k \gtrsim k_{\rm peak}/10$. 

We use this parametric model of USR and the associated numerical setup that 
computes scalar and tensor power spectra to compare against the datasets of CMB, 
galaxy surveys and Lyman-$\alpha$. We shall discuss the methods and results 
of our analyses in the following sections.


\section{Results} \label{sec:res}

In this paper, we aim to constrain the deviation of a class of USR models from 
the standard model of slow roll inflation followed by $\Lambda$CDM evolution. 
{\tt Planck} CMB temperature and polarization data puts strong constraints on such deviation at scales $10^{-3} \, {\rm Mpc^{-1} }\lesssim k \lesssim 0.2 \, {\rm Mpc^{-1}}$\footnote{The approximate smallest scale probed by {\tt Planck} data corresponds to  $k \simeq \ell/\eta_0$, where $\ell \simeq 2500$ and the conformal time at the present time $\eta_0 \simeq 11000 \, \rm Mpc$.}. Lyman-$\alpha$ data at intermediate redshifts ($2 \lesssim z \lesssim 4$) can probe scales comparable to the Jeans' scales in IGM ($k_{\rm J} \simeq 10\,\mpcinv$). We consider Planck data and other related data sets at comparable scales, along with high spectral-resolution Lyman-$\alpha$ data in our analysis. 

\subsection{CMB and galaxy survey data} \label{sec:cmbplus}

We consider the combined dataset of anisotropies in temperature and polarization 
of CMB from {\tt Planck 2018} and {\tt BICEP/KECK 2018 (BK18)}, along with
galaxy survey dataset from {\tt BOSS} and {\tt DES}~\cite{Planck:2019nip,BICEP:2021xfz,BOSS:2016wmc,Drlica-Wagner_2018}.
Given the sensitivity of the CMB data, we expect the strongest constraint on the 
model parameters from {\tt Planck} data. Hence, we first compare the model against the 
data of {\tt Planck 2018} and {\tt BK18}, and examine the constraints. 
We then extend the dataset to include {\tt BOSS} and {\tt DES} to study the 
improvement in the constraints.
We use the publicly available package called {\tt CosmoMC} to perform the 
comparison and Monte-Carlo sampling of the posterior distribution of 
parameters~\cite{Lewis:2002ah}.
We modify the {\tt CAMB} package within the setup of {\tt CosmoMC} to incorporate
the power spectra arising from the model of interest~\cite{Lewis:1999bs}.
We use a modified version of a package called {\tt PBS}\footnote{It is developed
by one of the authors and publicly available at 
\tt https://gitlab.com/ragavendrahv/pbs\,.}, which computes scalar and tensor power 
spectra for a given model of inflation, to implement the above mentioned
alteration to {\tt CAMB} in {\tt CosmoMC}.
Further, we use {\tt GetDist} to marginalize along various dimensions in the
parameter space and plot the resulting posterior distribution~\cite{Lewis:2019xzd}.

Since we are interested in studying the impact of enhanced  scalar power on small 
scales, we set the priors on the parameters of interest as $N_1=[25,35]$ and 
$\Delta N=[0,1]$\,.
The parameter $N_1$ determines the location of the enhancement of the scalar
power spectrum and the range of values that we have chosen corresponds to 
onset of enhancement over $k=[0.1,10^5]\,\mpcinv$\footnote{We should clarify we 
are not allowing for an epoch of USR at a much earlier time during inflation, 
which along with rescaling of the overall normalization, can suppress power over 
the largest scales probed by CMB, $k \simeq 10^{-4}\,\mpcinv$\,. 
In fact, it has been shown that such an epoch can mildly improve the fit to 
CMB~\cite{Qureshi:2016pjy,Ragavendra:2020old}\,. However, we are not 
allowing for this effect in our work and the priors on $N_1$ are chosen solely
to explore only the small-scale enhancement due to a late epoch of USR.}.

The parameter $\Delta N$ affects both the location and amount of enhancement of 
the spectrum. Our choice of the range corresponds to relative enhancement of 
about $[0,10^3]$ over the nearly scale-invariant amplitude.
As noted in Sec.~\ref{sec:USR}, we choose the range of $\Delta N$ such that
we do not violate the upper bound on $\ps(k) \lesssim 10^{-5}$ arrived at from 
the limit on spectral distortions obtained by {\tt FIRAS}~\cite{Mather:1993ij,Fixsen:1996nj,Chluba:2019kpb}.
The amplitude of the scalar and tensor power over the large scales prior to 
enhancement is determined by $H_{\rm i},\, \epsilon_{1_{\rm i}}$ and 
$\epsilon_{2_{\rm i}}$. Thus our inflationary model has five parameters and the
priors on these five model parameters along with those on other cosmological 
parameters are given in Tab.~\ref{tab:std-priors}.
\begin{table}
\begin{center}
\begin{tabular}{| c | c | c |}
\hline
Parameters & Minimum & Maximum \\ 
\hline
$\Omega_{\rm b}\,h^2$ & $5\times 10^{-3}$ & 0.10 \\ 
\hline
$\Omega_{\rm c}\,h^2$ & $10^{-3}$ & 0.99 \\  
\hline
$100\,\theta_{\rm MC}$ & 0.50 & 10.00 \\  
\hline
$\tau$ & $10^{-2}$ & 0.80\\
\hline
$\log_{10}(H_{\rm i}/\Mpl)$ & $-5.40$ & $-4.90$ \\ 
\hline
$\log_{10}(\epsilon_{1_{\rm i}})$ & $-4.25$ & $-3.40$ \\ 
\hline
$\log_{10}(\epsilon_{2_{\rm i}})$ & $-2.50$ & $-1.20$ \\ 
\hline
$N_1$ & 25.00 & 35.00 \\ 
\hline
$\Delta N$ & 0.00 & 1.00 \\ 
\hline
\end{tabular}
\caption{The priors on the cosmological parameters and those on our inflationary
model parameters that are used while comparing the models against datasets of 
{\tt Planck 2018, BK18, BOSS} and {\tt DES} are listed. We have varied the 
inflationary parameters determining the power spectrum over large scale in 
logarithmic scale to explore a wide region in the parameter space.}
\label{tab:std-priors}
\end{center}
\end{table}

\subsubsection{Five parameter model}

We first compare the model against the CMB datasets by using the likelihoods of 
{\tt Planck2018} and {\tt BK18}.
We sample the posterior distribution and obtain the constraints on the marginalized
distribution of the parameters of interest. 
We find that the four cosmological parameters, namely 
$\Omega_{\rm c}h^2$, $\Omega_{\rm b}h^2$, $\theta_{\rm MC}$ and $\tau$ are 
constrained around the same value as they are in PL model.
Besides, the constraints on $H_{\rm i},\,\epsilon_{1_{\rm i}}$ and $\epsilon_{2_{\rm i}}$, 
are such that they lead to a nearly-scale invariant behavior over the CMB scales. 

As to the parameters of USR, $N_1$ is bounded from below. 
To quantify this lower limit, we use {\tt GetDist} which deals with one-sided 
distributions and sharp boundaries in the distribution using a linear boundary 
kernel estimator and suitable normalization~\cite{Lewis:2019xzd}. 
We obtain that $27.26 < N_1 < 33.29$ at the level of $1$-$\sigma$ and $N_1 > 26.61$ 
at the level of $2$-$\sigma$. On the other hand, $\Delta N$ is unconstrained with 
a mild preference towards $\Delta N = 0$.
To avoid clutter, we have not plotted the contours of the marginalized posterior
distribution of this analysis and plot the distribution only for the following case.

\begin{figure}[H]
\centering
\includegraphics[scale=0.45]{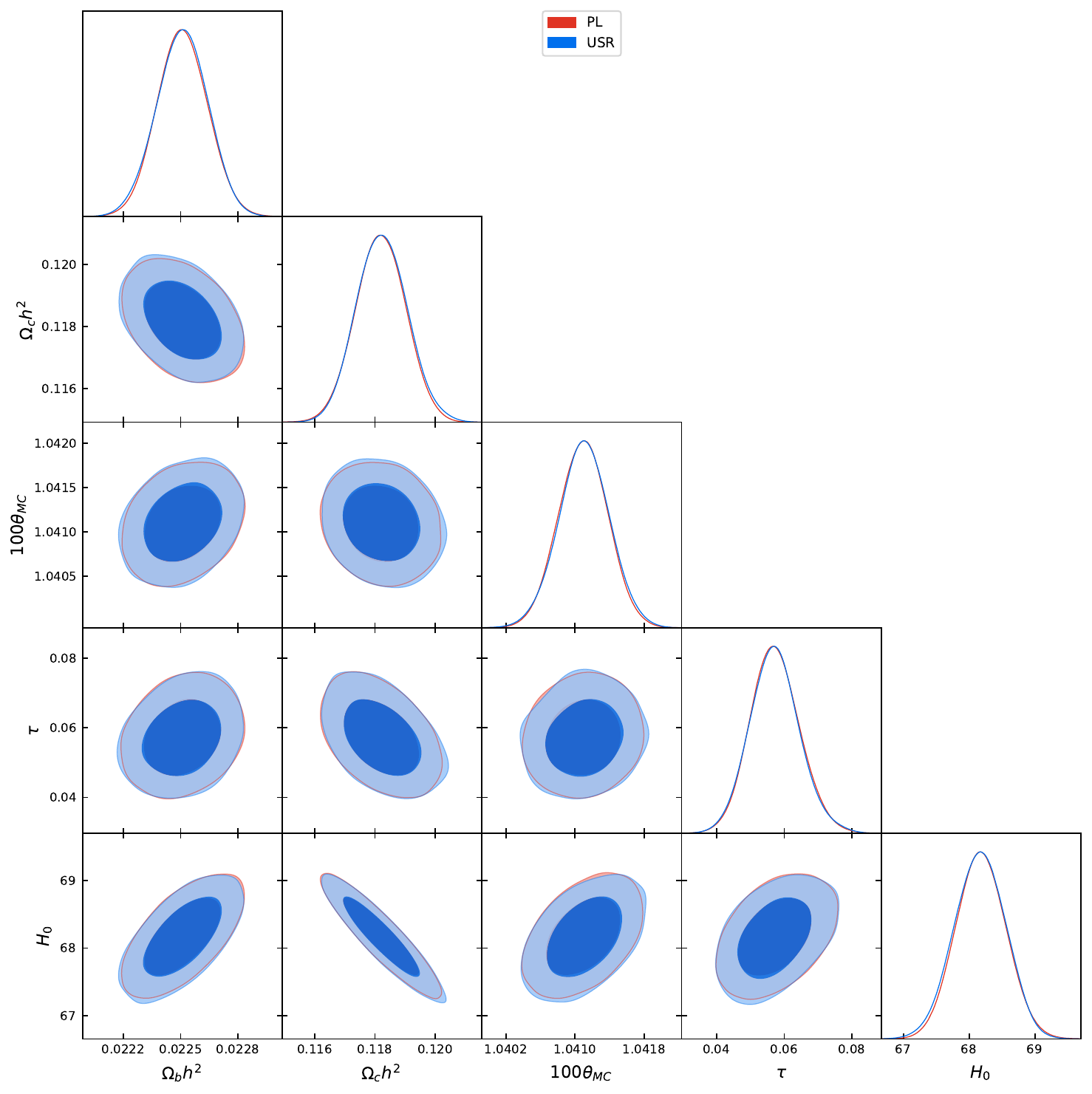}
\includegraphics[scale=0.25]{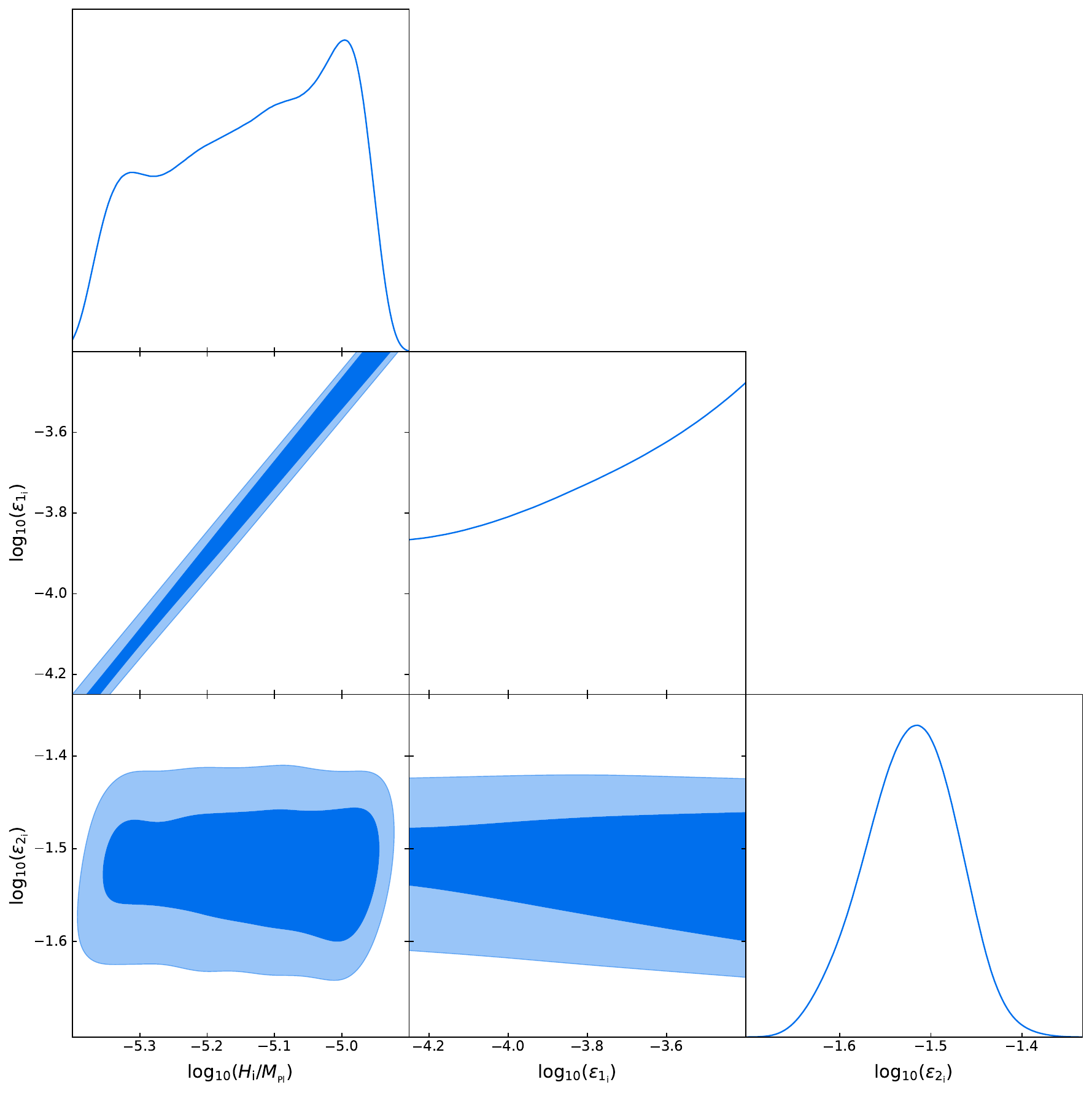}
\includegraphics[scale=0.25]{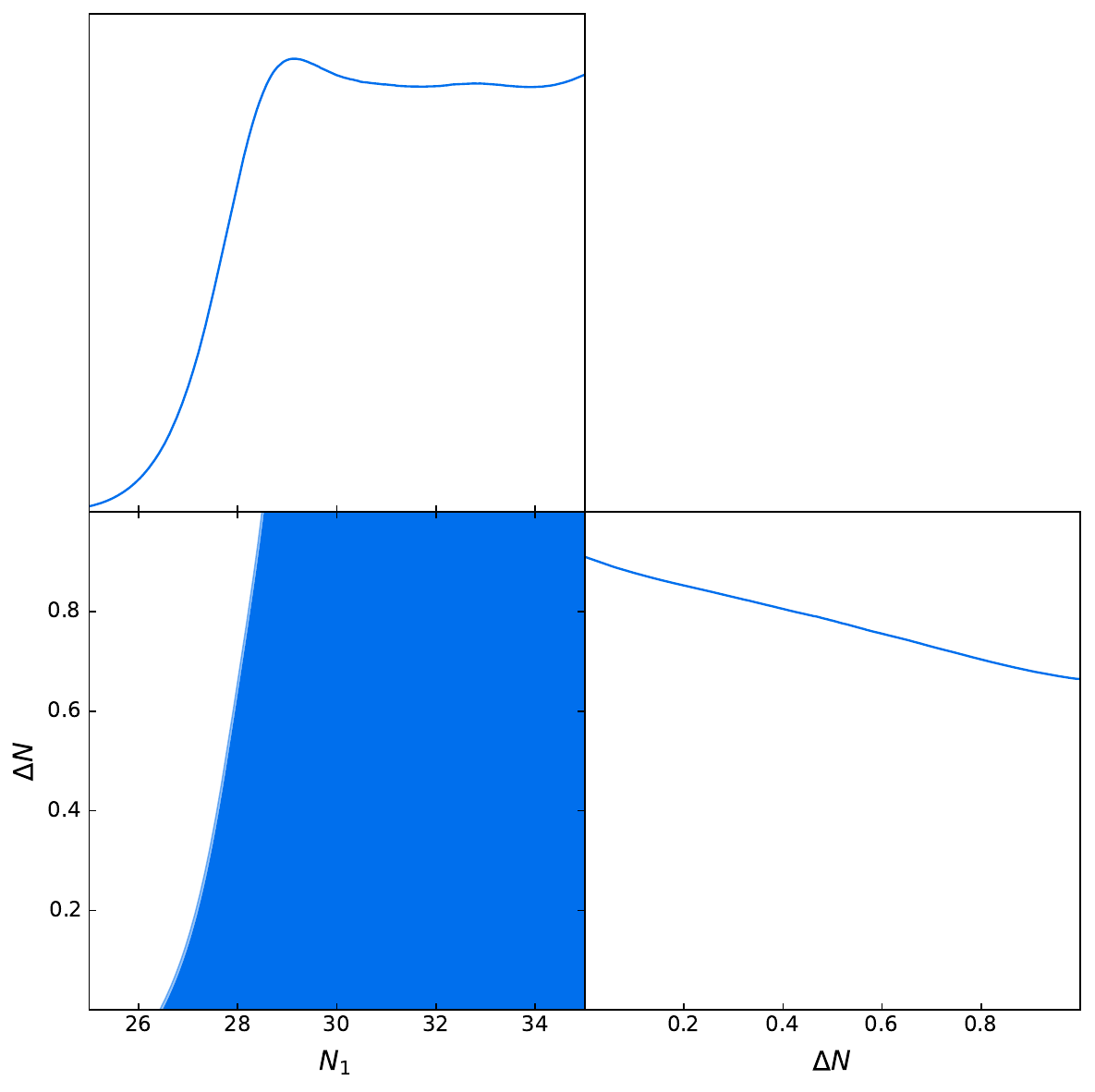}
\vskip -0.1 in
\caption{The $1$-$\sigma$ and $2$-$\sigma$ contours of the posterior distribution of the 
parameters of USR model when compared against CMB ({\tt Planck2018+BK18}) 
and galaxy survey ({\tt BAO+DES}) data are presented here. 
The standard cosmological parameters are presented (on the top panel). 
Their distributions are the essentially the same between PL and USR models.
The three parameters that determine the spectrum on large scales, 
$\log_{10}(H_{\rm i}/\Mpl)$, $\log_{10}(\epsilon_{1_{\rm i}})$ and 
$\log_{10}(\epsilon_{2_{\rm i}})$, are presented (on bottom left) along with the 
parameters dictating USR epoch, $N_1$ and $\Delta N$ (on bottom right).}
\label{fig:USR_5_params_CMB+gal}
\end{figure}

Next, we extend the dataset to include galaxy surveys, i.e. {\tt BOSS} and {\tt DES} 
likelihoods and repeat the above exercise. 
We present the marginalized contours of the posterior distribution of this analysis
in Fig.~\ref{fig:USR_5_params_CMB+gal}. We once again observe that the parameters 
such as $\Omega_{\rm c}h^2$, $\Omega_{\rm b}h^2$ get constrained around the same 
values as in the PL model. 
The onset of USR is constrained as $N_1 > 29.73$ at $1$-$\sigma$ level and 
$N_1 > 27.53$ at $2$-$\sigma$ level.
$\Delta N$ is again unconstrained, though there is mild preference towards 
$\Delta N=0$. 

The constraints on the parameters $H_{\rm i},\,\epsilon_{1_{\rm i}}$ and 
$\epsilon_{2_{\rm i}}$ are similar to those obtained when comparing against CMB alone.
The marginalized contours of $H_{\rm i},\, \epsilon_{1_{\rm i}}$ and $\epsilon_{2_{\rm i}}$
can be understood through their relations to the conventional parameters that
describe the spectrum over large scales, namely the amplitude of scalar power at
pivot scale $A_{\rm s}$, the spectral index of the scalar power spectrum $\ns$ and 
the tensor-to-scalar ratio $r$. These parameters describe the shape and amplitude
of the power spectra in the PL model~\cite{Planck:2018jri}.
Under slow roll approximation, which is valid over the initial slow roll regime of 
our model, we can relate our model parameters $H_{\rm i},\,\epsilon_{1_{\rm i}}$ 
and $\epsilon_{2_{\rm i}}$ to these PL parameters as~(see, for example~\cite{Kinney:2009vz,Baumann:2009ds,Sriramkumar:2009kg})
\bea 
A_{\rm s} & \simeq & \f{H^2(N_{\rm p})}{8\pi^2\Mpl^2\epsilon_1(N_{\rm p})}\,,\\ 
& \simeq & \f{H^2_{\rm i}}{8\pi^2\Mpl^2\epsilon_{1_{\rm i}}}{\rm e}^{-\epsilon_{2_{\rm i}}N_{\rm p}}\,, \\
\ns & \simeq & 1 - 2\epsilon_1(N_{\rm p}) - \epsilon_2(N_{\rm p})\,, \\
& \simeq & 1 - 2\epsilon_{1_{\rm i}}{\rm e}^{\epsilon_{2_{\rm i}}N_{\rm p}} - \epsilon_{2_{\rm i}}\,, \\
r & \simeq & 16\epsilon_1(N_{\rm p})\,, \\
& \simeq & 16\epsilon_{1_{\rm i}}{\rm e}^{\epsilon_{2_{\rm i}}N_{\rm p}}\,.
\eea
The above quantities are evaluated at $N_{\rm p}$, when the pivot scale $k_{\rm p}$ 
exits the Hubble radius, which we have set to be $N_{\rm p}=22$ as mentioned before.
We have used the fact that the second slow-roll parameter is a constant 
$\epsilon_{2_{\rm i}}$ during the initial slow roll regime and so
$\epsilon_1(N_{\rm p}) = \epsilon_{1_{\rm i}}\exp(-\epsilon_{2_{\rm i}}N_{\rm p})$
[cf. Eq.~\eqref{eq:eps1-parametric}].
Also, the Hubble parameter $H(N_{\rm p})$ can be evaluated by evolving it from 
its initial value $H_{\rm i}$, as dictated by Eq.~\eqref{eq:H-evol}. 
However, the value of $\epsilon_1(N_{\rm p})$ is so small that it makes little 
difference to $H(N)$ during the evolution. 
Hence, we have assumed $H(N_{\rm p}) \simeq H_{\rm i}$.

\begin{figure}[H]
\centering
\includegraphics[scale=0.3]{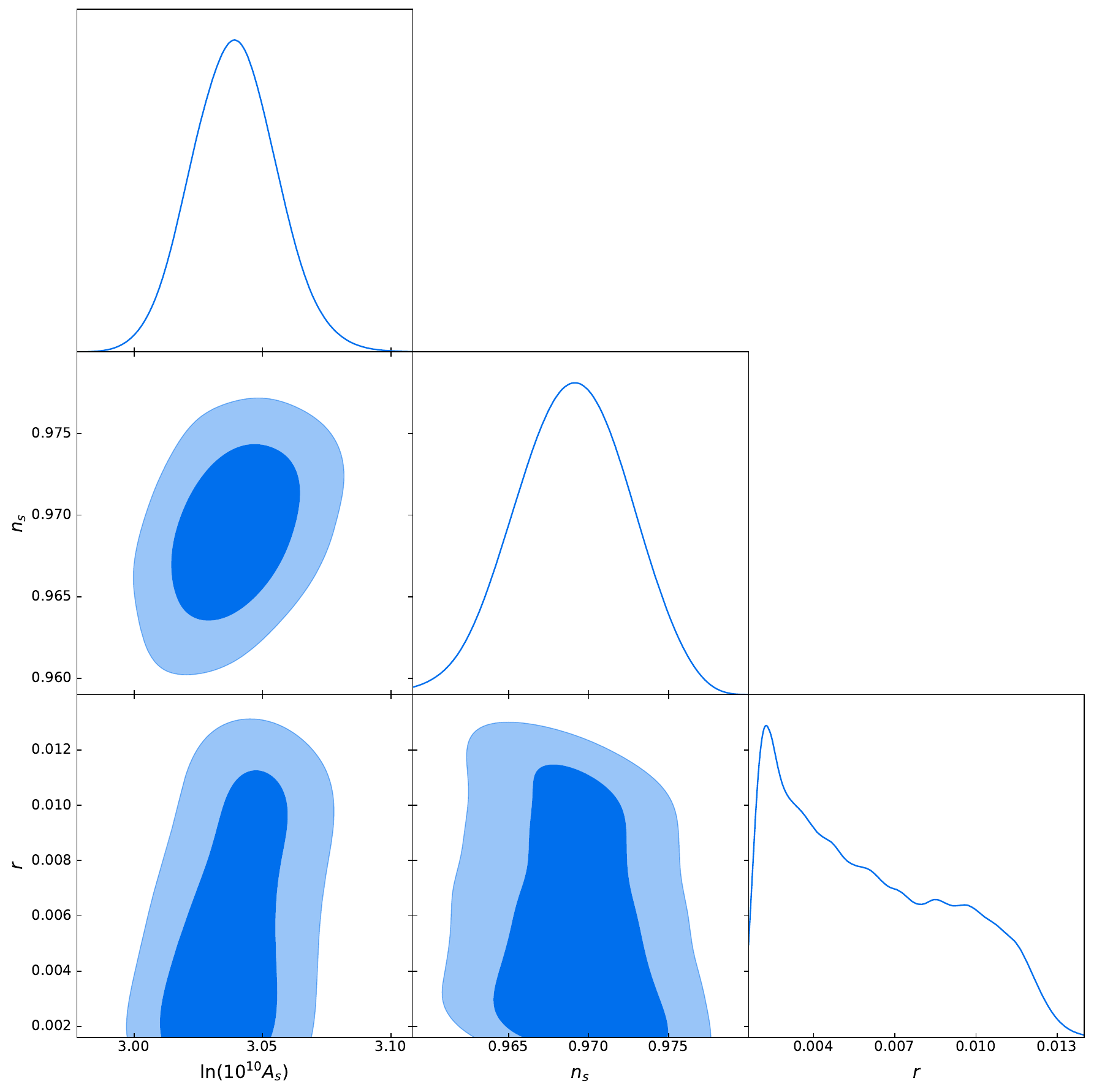}
\vskip -0.1 in
\caption{The $1$-$\sigma$ and $2$-$\sigma$ contours of the posterior distribution of the 
parameters $A_{\rm s}, \ns$ and $r$ derived from $H_{\rm i},\epsilon_{1_{\rm i}}$ 
and $\epsilon_{2_{\rm i}}$, using slow-roll approximation over the regime before
the onset of USR, are presented here. 
The apparent closing of contour on $r$ with a lower bound is due to the lower
limit of the range of prior on $\epsilon_{1_{\rm i}}$.}
\label{fig:As-ns-r-derived-CMB+gal}
\end{figure}

However, we should note that we do not directly vary $A_{\rm s},\,\ns$ or $r$ while 
comparing our model against data. 
We vary the model parameters $H_{\rm i},\,\epsilon_{1_{\rm i}}$ and $\epsilon_{2_{\rm i}}$,
along with $N_1$ and $\Delta N$, and numerically evolve the background and perturbations 
for each given value of them, when comparing against data.
Hence, $A_{\rm s},\,\ns$ or $r$ are parameters derived from the model parameters
$H_{\rm i},\,\epsilon_{1_{\rm i}}$ and $\epsilon_{2_{\rm i}}$
through the relations given above. We present the contours of these derived
parameters in Fig.~\ref{fig:As-ns-r-derived-CMB+gal}.
The amplitude $A_{\rm s}$ being constrained corresponds to $H_{\rm i}$ being
strongly correlated with $\epsilon_{1_{\rm i}}$. 
Note that $A_{\rm s},\,\ns$ and $r$ are exponentially sensitive to 
$\epsilon_{2_{\rm i}}$ due to the time evolution of $\epsilon_1(N)$.
Therefore, constraints on these three parameters, especially on $\ns$, correspond 
to the strong constraint on $\epsilon_{2_{\rm i}}$.
The tensor-to-scalar ratio $r$ being bound from above mainly translates to the
unconstrained behavior of $\epsilon_{1_{\rm i}}$, as the range of prior on
$\epsilon_{1_{\rm i}}$ is well within the bound on $r$\,.
Besides, the parameter $r$ is also indirectly related to $\ns$, through its 
dependence on $\epsilon_{2_{\rm i}}$\,.
Thus the closing of contour on $r$ with a lower bound is due to a combination
of the lower limit of the prior on $\epsilon_{1_{\rm i}}$ and the tight
constraint on $\epsilon_{2_{\rm i}}$\,.
It is worth noting that, if it was parametrized as $r \simeq 16\epsilon_{1_{\rm i}}$, 
neglecting the time evolution of $\epsilon_1$, we would have obtained only an 
upper bound and no lower bound on $r$.

\subsubsection{Two parameter model}

An important takeaway from the above analyses is that the constraints on the
parameters determining the first slow roll phase are largely independent of the
USR parameters. This is because the slow-roll and USR dynamics are fairly
independent of each other in our model and they affect different ranges of scales. 
To illustrate this point more clearly, we compare our model against the dataset 
of CMB and galaxy surveys, by varying just the USR parameters $N_1$ and $\Delta N$, 
while fixing the other three parameters $H_{\rm i},\epsilon_{1_{\rm i}}$ and 
$\epsilon_{2_{\rm i}}$ at values well within their constraints as follows:
$\log_{10}(H_{\rm i}/\Mpl)=-5.14,\,\log_{10}(\epsilon_{1_{\rm i}})=-3.82$
and $\log_{10}(\epsilon_{2_{\rm i}})=-1.47$\,.
We find that the constraints on $N_1$ and $\Delta N$ are very similar to those
obtained in the previous analysis when we varied all the five model parameters.
We obtain $N_1 > 29.85$ at 1-$\sigma$ level and $N_1 > 27.56$ at 2-$\sigma$ level, 
matching the bounds gotten earlier. $\Delta N$ is unconstrained as before with a 
mild preference towards zero.

However, the fixing of the inflationary three parameters reduces the computational 
time taken for sampling the posterior distribution, while still arriving at the correct 
constraints on USR parameters. This confirms the separation of scales between slow roll 
and USR dynamics. One can study the USR phase affecting the small scales while not 
altering much of the slow roll dynamics affecting the large scales and arrive at the 
same conclusion, as while considering the complete dynamics of our model.

We should also caution that there may be certain potentials realizing USR dynamics, where
the parameters dictating USR may also affect the shape of spectra over large scales. 
Such potentials may typically lead to strong running of $\ns$ as mentioned earlier.
These effects can alter the fit to CMB and hence push the onset of USR further to a
later time during inflation. Thus, our bounds on $N_1$ are relatively conservative 
in this regard.

The lower bound obtained on $N_1$ is an important result of these analyses so far. 
The bound at the level of 2-$\sigma$ of $N_1 > 26.61$ from CMB is improved by about an 
e-fold as $N_1 > 27.56$ with addition of galaxy surveys to the dataset.
\color{black}
It translates to an upper bound on $\bar N_1$, the earliest onset of USR from 
the end of inflation. The bound of $N_1 > 27$ is simply $\bar N_1 < 45$,
implying that USR can occur only in the last $45\,$ e-folds of inflation.
It essentially indicates the data's preference for the nearly scale-invariant 
spectrum up to scales $k\lesssim1\,\mpcinv$ and avoidance of any deviation such 
as a dip or rise over these scales.
On the other hand, the indifference of data towards the range of $\Delta N$
suggests that an epoch of USR lasting for as long as $1$ e-fold beyond $N_1=27$ 
and hence a relative enhancement as large as $10^3$ past $k=1\,\mpcinv$ does 
not affect the fit of the model over large scales.
We note that this scale is comparable to but smaller than our expectation based on 
the relation $k\eta_0 = \ell$, which follows from the property of spherical 
Bessel function. 
We note that for a given angular scale $\ell$, the contribution from a linear 
scale $k < \ell/\eta_0$ falls very sharply. However, it only falls as 
$1/(k\eta_0)$ for $k > \ell/\eta_0$ (for details see e.g.~\cite{dodelson}). 
This means CMB data is sensitive to much larger values of $k$ for a fixed $\ell$ 
and $\eta_0$ as our results show. 

\FloatBarrier
\subsection{Lyman-$\alpha$ forest} \label{sec:lyalpha}

Lyman-$\alpha$ forest is the dense, ragged forest of lines seen in the spectra of 
the background quasars. These lines correspond to regions in the Inter-Galactic 
Medium (IGM) with HI column number density lying between the range:
$N_{\HI} = 10^{12} \, \hbox{--} \, 10^{15}$ atoms/cm$^{2}$. Hydrodynamical 
simulations identify these regions with underlying 
matter densities in the mildly non-linear range i.e. 
$\delta \leq 10$. It is known that Lyman-$\alpha$ forest can probe density perturbations on scales 
as small as the thermal Jeans' scale of IGM at intermediate redshifts ($k \simeq 10 \, \rm Mpc^{-1}$). 
Based on hydrodynamical simulation of Lyman-$\alpha$ forests, Murgia et~al. \cite{Murgia:2017lwo} put constraints on a large range of models with matter  power deficit as compared to the $\Lambda$CDM model. Most of the models
we consider correspond to excess power as compared to the $\Lambda$CDM model. While it is tempting to 
use the results of \cite{Murgia:2017lwo} to constrain our models, the relevant cosmological 
observable in this case, the flux deficit, is a non-linear function of matter power and therefore 
we recompute these constraints based on a semi-analytic approach \cite{bi1997evolution,hui1997equation,choudhury2001semianalytic,pandey2012probing,sarkar2021using}. 

The semi-analytic modelling of Lyman-$\alpha$ has been used to constrain cosmological models which either have excess or reduced matter power in \cite{pandey2012probing,sarkar2021using}. In this approach,
the  observed redshift evolution of effective optical depth of Lyman-$\alpha$ forest, based on high spectral-resolution data \cite{faucher2008direct}, is used. 

To constrain our model against the Lyman-$\alpha$ data, we define and use the 
relative difference between the 1D baryonic power spectra of the models of interest 
$\barpow1d(k)$ and the standard $\Lambda$CDM evolution following 
slow-roll inflation leading to a power-law primordial power spectrum $\barpow1d(k)\vert_{_{\rm PL}}$. The 1D baryonic power spectrum can be defined in terms of the 3D baryonic power 
spectrum $P^{(3)}_b(k,z)$ as (e.g. \cite{bi1997evolution,hui1997equation} for details): 
\begin{equation}
P^{{(1)}}_b (k, z) = \frac{1}{2 \pi} \, \int_{|k|}^{\infty} \, k^{'} \, P^{(3)}_b(k^{'}, z) \, dk^{'} 
\label{eq:pk1db}
\end{equation}
with 
\begin{equation}
P^{(3)}_b(k,z) = \frac{P^{(3)}_m(k)}{\left[ \, 1  + (k/k_J)^2 \, \right]^2}  \, \, ,
\label{eq:Pb3D}
\end{equation}
and the Jeans wavenumber $k_{\rm J}$ is:
\begin{equation}
k_{\rm J} = H_0\left[ \frac{2 \gamma k_B T_0}{3 \mu m_p 
\Omega_m (1+z)} \right]^{-1/2} \, \, , 
\label{eq:jeans}
\end{equation}
where the parameters 
have their usual meaning, with  $\mu = 0.6$. For the standard choice of parameters, we have $k_{\rm J} \simeq 8.5 \, h \, {\rm Mpc}^{-1}$  at the redshift $z = 3$. 

We further average the relative difference $\pmrd(k)$  between 1D baryonic power spectra over the range of  $k$. This allows us to define  $\Delta^{1\rm D}_{\rm m}$ as follows:
\begin{eqnarray}
\pmrd(k,z) &=& \f{\barpow1d(k,z)-\barpow1d(k,z)\vert_{_{\rm PL}}}{\barpow1d(k,z)\vert_{_{\rm PL}}}\,,  \nonumber \\
\Delta^{1\rm D}_{\rm m}(z) &=& \f{1}{\int \d \ln k}\int \d \ln k\,\pmrd(k,z)\,.
\label{eq:avg-rel-diff}
\end{eqnarray}
We note that $\Delta^{1\rm D}_{\rm m}(z)$ is a function of redshift owing to the evolution of Jeans scale (Eq.~(\ref{eq:jeans})). To parameterize the relative difference, $\Delta^{1\rm D}_{\rm m}(z)$ is  computed at the redshift of $z_\ast=2.2$. The choice of $z_\ast$ has no impact on our main results. 
The range of 
integration over $k$ is taken to be $[10^{-5},\,10^3]\,\mpcinv$
to arrive at $\Delta^{1\rm D}_{\rm m}(z_\ast)$\,.
Our approach is similar to Murgia et~al. (2017) \cite{Murgia:2017lwo} where the deviation from
the standard 1D power spectrum is quantified and compared against data. We differ from the measure
used by \cite{Murgia:2017lwo} in using baryonic power spectra and using a larger range of $k$ to define
the relative difference between the standard and modified models. 
We use a semi-analytic model for probing the thermal, dynamical and 
ionization state of the Lyman-$\alpha$ clouds at higher redshifts using the following parameters:
$J_0$ (the intensity of hydrogen-ionizing photons), $T_0$ (the temperature of the IGM), $\gamma$ (the equation of state of the Lyman-$\alpha$ clouds) (for details see \cite{sarkar2021using} and references therein). Based on observational data and hydrodynamical simulations, these parameters are varied in
the following range: $7000 \leq T_0 \leq 15000 {\K}$, $0.7 \times 10^{-12} \, {\rm sec^{-1}} <J_0 < 2 \times 10^{-12} \, \rm sec^{-1}$, and  $0.9 \leq \gamma \leq 2.2$ (for details see \cite{sarkar2021using}). 
In addition, two cosmological parameters, $N_1$ and $\Delta N$, denoting the 
modified small scale power owing to USR phase, are included in the analysis. 

We have simulated the 1D density and velocity field using lognormal distribution  in the redshift interval $2 \le z \le 4.2$ at uniform
redshift intervals of $\Delta z = 0.1$ for comparison with data. The Lyman-$\alpha$ effective optical depth $\tau_{\rm eff} (z)$ is defined as: $\tau_{\rm eff} (z) = - {\rm log}\left[ \, \langle \, {\rm exp}(-\tau) \, \rangle \, \right]$. From the standpoint of observation, 
the ensemble average (represented by the angular brackets $\langle \hbox{...} \rangle$) 
is substituted by the average over the number of individual Lyman-$\alpha$ clouds at a particular redshift:
$\tau_{\rm eff} (z) = - {\rm log} \, \left[ \, \sum_{i} \, {\rm exp}(-\tau_i) / N \, \right]$, 
where $N$ is the number of the simulated  Lyman-$\alpha$ clouds at a  redshift $z$, 
and $\tau_i$ denotes
the optical depth of the $i$-th Lyman-$\alpha$ cloud at that redshift. 

\begin{figure}
\centering
\includegraphics[scale=0.5]{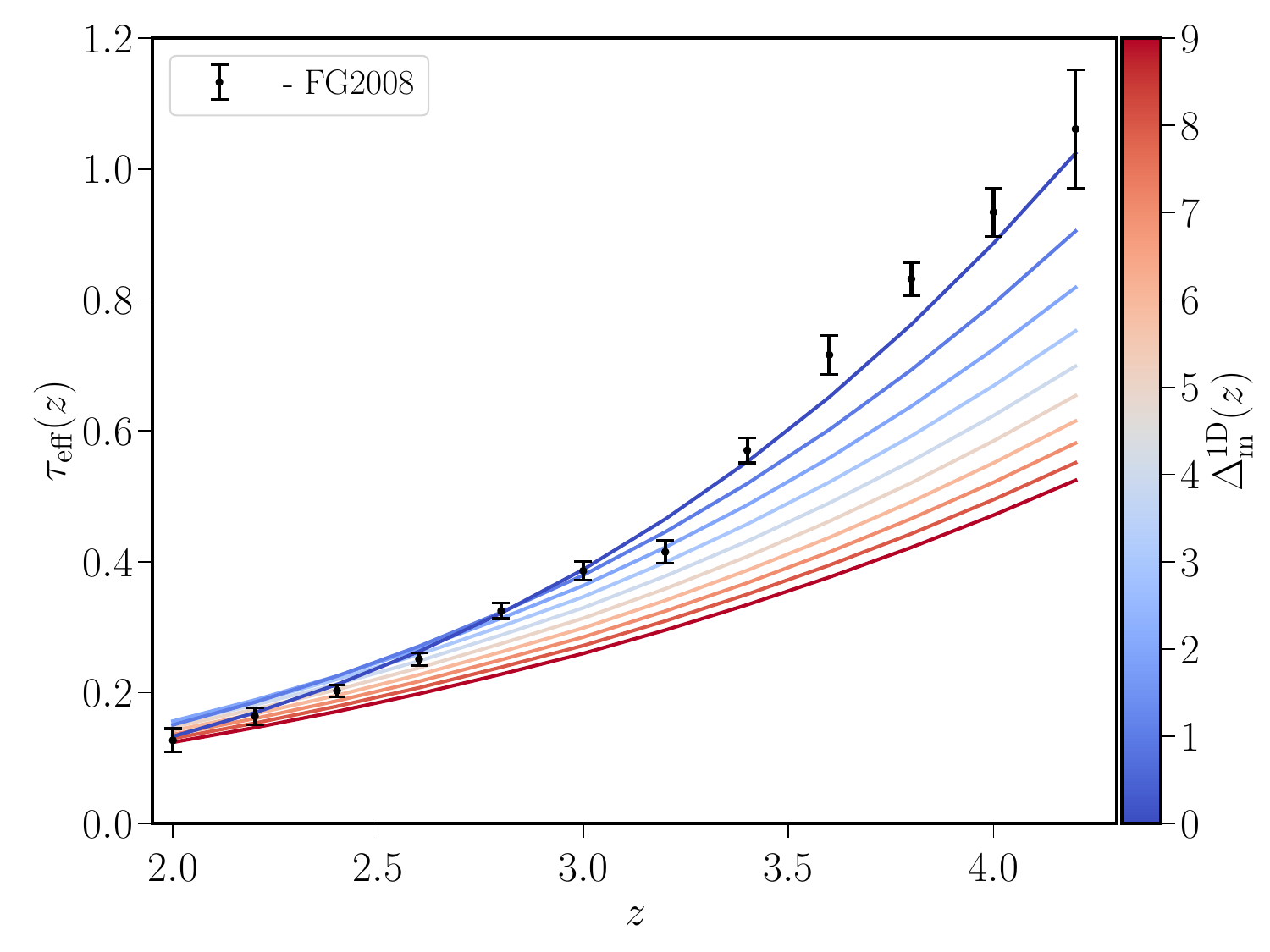}
\vskip -0.1in
\caption{The evolution of $\tau_{\rm eff}$ is shown based on our model predictions 
along with the high spectral-resolution data  (in black dots along with error bars, 
FG2008)~\cite{faucher2008direct}. Different curves (from top to bottom) correspond 
to increasing  average relative difference of 1D power spectrum 
$\Delta^{\rm 1D}_{\rm m}(z)$ [cf.~Eq.~\eqref{eq:avg-rel-diff}] (in shades of blue 
to red). In simulating the evolution of  $\tau_{\rm eff} (z)$ for different values of 
$\Delta^{\rm 1D}_{\rm m}(z)$, the Lyman-$\alpha$ parameters 
have been held fixed to the following values: 
$J_0  = 10^{-12} \, {\rm s}^{-1}$, $T_0 = 2.3 \times 10^4 \, {\rm K}$
and $\gamma = 0.7$.}
\label{fig:taucompare}
\end{figure}
In  Fig.~\ref{fig:taucompare}, we plot the evolution of $\tau_{\rm eff}$ along with the data. 
To understand the evolution of $\tau_{\rm eff}$, we first discuss two extreme cases. If $\tau_i \ll 1$ for all clouds, then $\tau_{\rm eff} \propto (1+z)^{4.5}$, which is the expected evolution of the Gunn-Peterson optical depth for the background universe. In the other extreme case, if $\tau_i \gg 1$, then $\tau_{\rm eff}$ is independent of $\tau_i$ and hence of redshift. The observed behaviour lies between these two
extremes which shows that many clouds are making transition from optical thin to optically thick as
redshift increases. ($\tau_i = N_{\rm HI} \sigma_\alpha$, where $\sigma_\alpha$ is the cross-section 
of Lyman-alpha scattering. The line center cross-section $\simeq 5 \times 10^{-14} \, \rm cm^{-2}$ for 
$T_0 \simeq 10^4$.) The data fits the $\Lambda$CDM model well  and can be used to constrain models with deficit or excess power~\cite{pandey2012probing}. Fig.~\ref{fig:taucompare} shows the redshift evolution of $\tau_{\rm eff}$ get flatter as the matter power is increased. This is expected as a larger 
fraction of Lyman-$\alpha$ clouds become optically thick at higher redshift as the matter power is increased. This demonstrates that the slope of the $\tau_{\rm eff}$ evolution is a robust probe of 
matter power (for further details see \cite{pandey2012probing,sarkar2021using} for cases analysing both deficit and excess of power). However, the effective optical depth is also affected by parameters
used to model the ionization and thermal state of clouds. In particular, the increase in matter power
is degenerate with ionizing intensity $J_0$ and partially degenerate with $T_0$ and $\gamma$ (see \cite{sarkar2021using} for details). Therefore, the magnitude of the change  of matter power 
that can be be probed by the data depends on the prior information on these parameters. We discuss the priors on these parameters above. These correspond to  the widest
possible priors on these parameters from the literature. Based on our analysis of the simulated data,
we find that  these priors constrains the change in  the 1D matter power spectrum to be smaller than roughly a factor of five. For narrower priors, the constraints are stronger. If the results of \cite{Murgia:2017lwo} are directly applied to our case, the change in the matter power is constrained to be less than approximately  50\%. 

\begin{figure}[!t]
\centering
\includegraphics[scale=0.5]{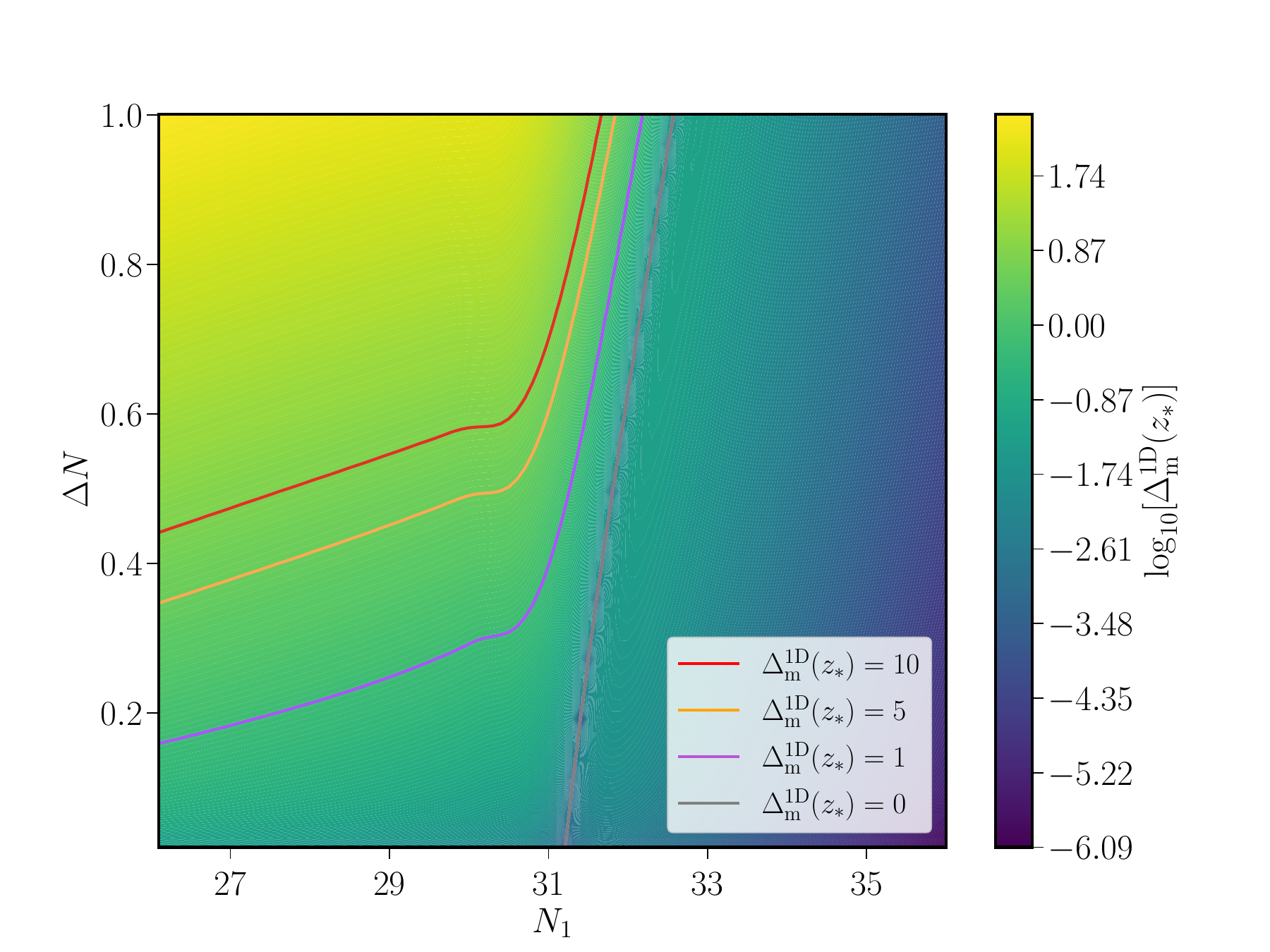}
\vskip -0.1in
\caption{We present the density plot of the average relative difference of the 
1D matter power spectrum of the USR model with respect to the standard power law
power spectrum $\Delta^{1\rm D}_{\rm m}(z_\ast)$ [Eq.~(\ref{eq:avg-rel-diff})], as 
a function of $N_1$ and $\Delta N$ (in shades of blue to green). We have also 
highlighted the contours corresponding to $\Delta^{1\rm D}_{\rm m}(z_\ast)=10, 5, 1$ and $0$ 
(with lines in red, orange, violet and grey respectively). 
In the region to the right of the line corresponding to 
$\Delta^{1\rm D}_{\rm m}(z_\ast)=0$, the parameters lead to 
$\Delta^{1\rm D}_{\rm m}(z_\ast)$ being small and negative in value.}
\label{fig:rel-diff-contours}
\end{figure}
\begin{figure}[!t]
\centering
\includegraphics[scale=0.3]{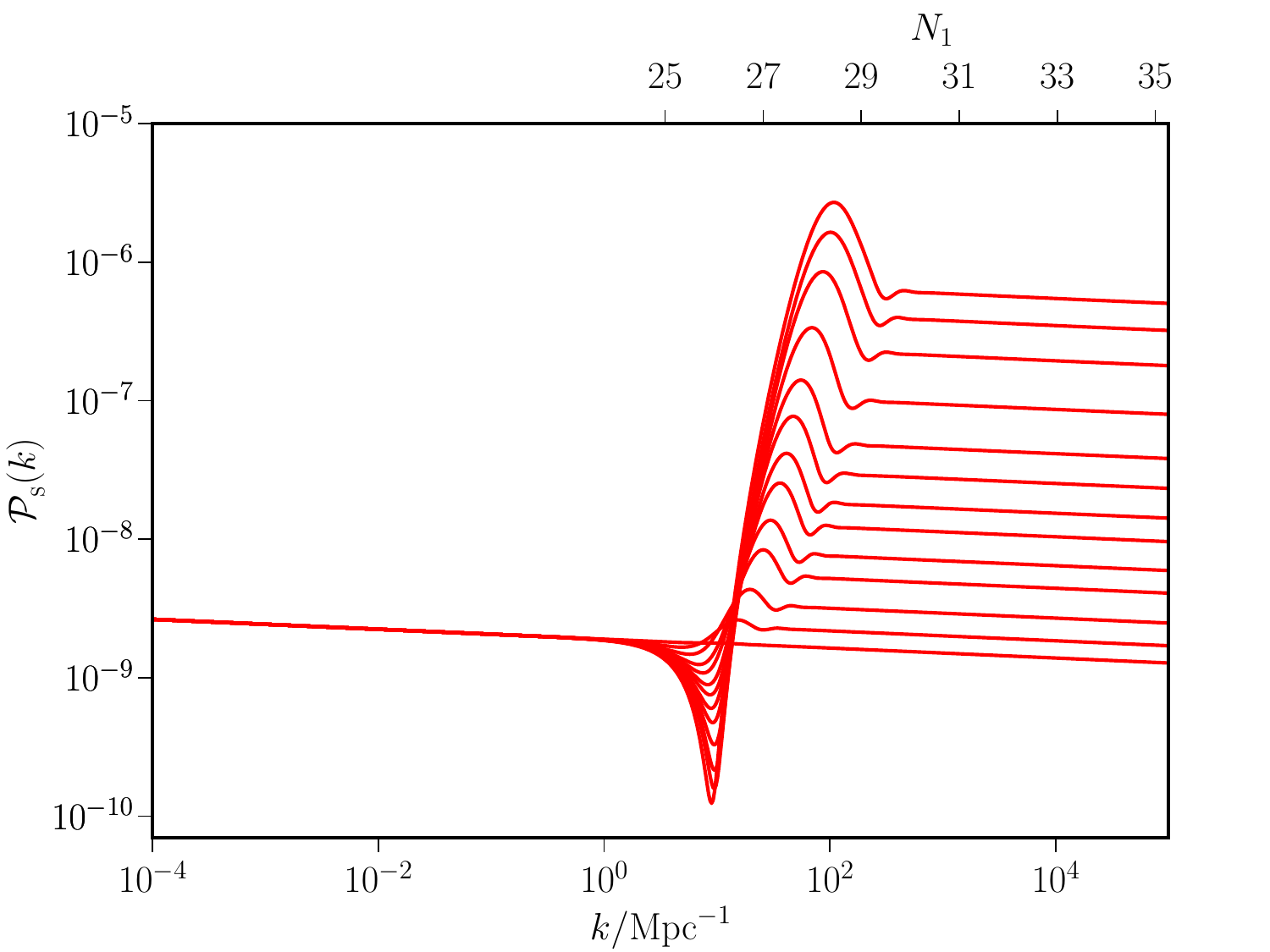}
\includegraphics[scale=0.3]{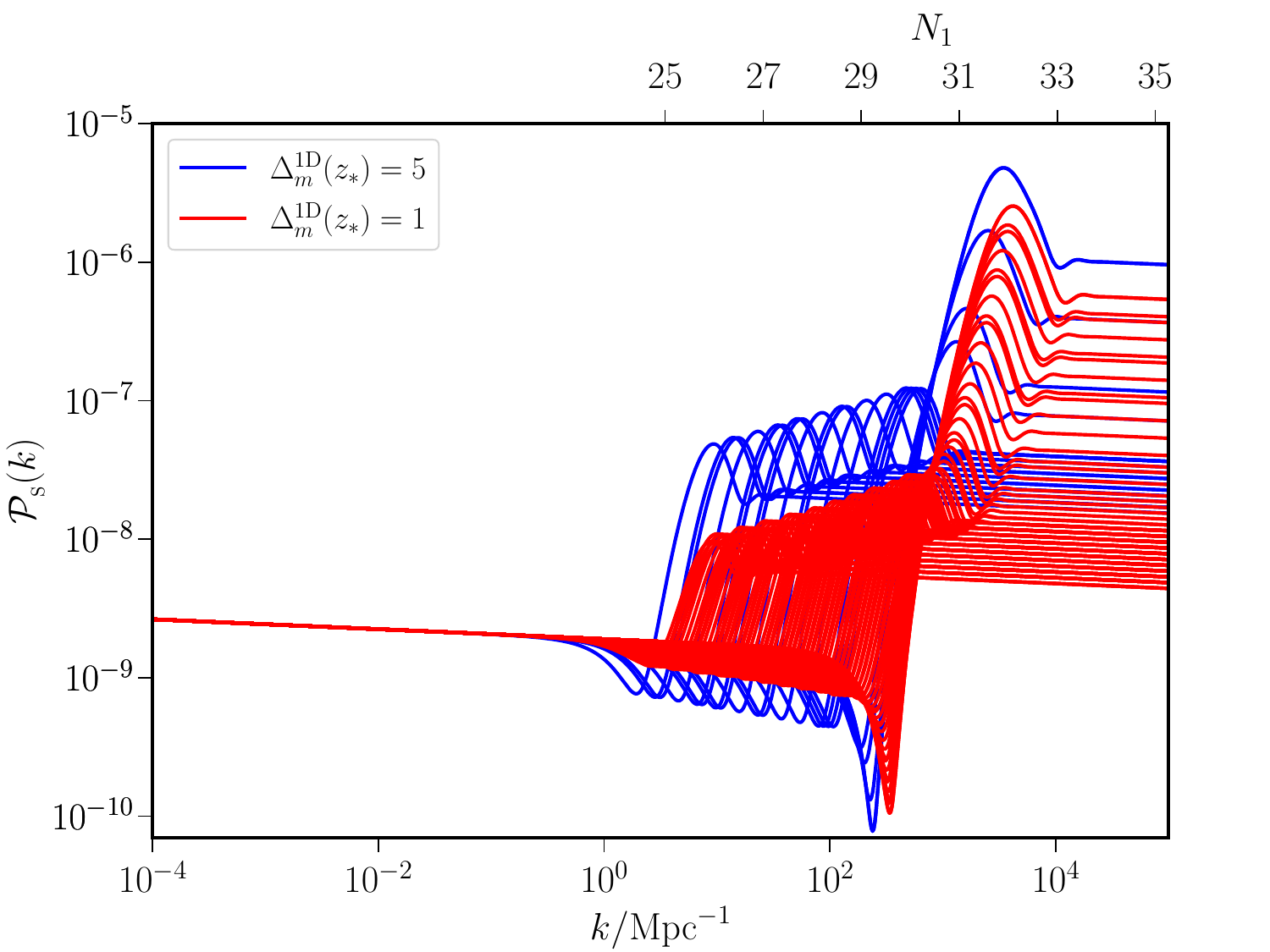}
\vskip -0.1in
\caption{Left Panel:~The scalar power spectra based on $2$-$\sigma$ bound 
on parameters $N_1$ and $\Delta N$ from CMB and galaxy surveys
({\tt Planck2018+BK18+BOSS+DES}) are shown
(cf.~Fig.~\ref{fig:USR_5_params_CMB+gal}). 
Right Panel:~The scalar power spectra  corresponding to constraints from 
Lyman-$\alpha$ data are displayed. The scalar power spectra leading to 
$\Delta^{1\rm D}_{\rm m}(z_\ast)=1$ and $\Delta^{1\rm D}_{\rm m}(z_\ast)=5$ are
presented (in red and blue respectively). These spectra are obtained from the 
values of $N_1$ and $\Delta N$ along the respective contours in 
Fig.~\ref{fig:rel-diff-contours}.
The upper x-axis displays  $N_1$ corresponding to the 
location of the peak, $k_{\rm peak}$, in the spectrum. It is obtained using the relation: 
$N_1 \simeq 25 + \ln(k_{\rm peak}/\mpcinv) - \ln(3.48)$ [cf.~Eq.~(\ref{eq:n1krel1})].}
\label{fig:Ps-rel-diffs}
\end{figure}

In Fig.~\ref{fig:rel-diff-contours}, we present the constraints from Lyman-$\alpha$
data as a color gradient map with contours based on the deviation measure, 
$\Delta^{1\rm D}_{\rm m}(z_\ast)$, in the  $N_1$--$\Delta N$ plane. 
The contours denoting the values of $\Delta^{1\rm D}_{\rm m}(z_\ast)=1, 5$ and $10$ are 
displayed. The region above these contours is ruled out. To understand these constraints, 
we show models corresponding to two values of $\Delta^{1\rm D}_{\rm m}(z_\ast)$ in
Fig.~\ref{fig:Ps-rel-diffs}.
In Fig.~\ref{fig:rel-diff-contours} we also plot a curve corresponding to 
$\Delta^{1\rm D}_{\rm m}(z_\ast) = 0$. 
This curve demarcates the region of positive deviation from negative deviation. 
The negative deviation is caused by 1-D spectrum getting suppressed by the dip in 
matter power and the subsequent rise suppressed by the cutoff at Jean's scale 
[cf.~Eq.~\eqref{eq:Pb3D}]. However, all the spectra that we consider 
(Fig.~\ref{fig:Ps-rel-diffs}) yield only a small negative deviation, 
$|\Delta^{1\rm D}_{\rm m}(z_\ast)| \le 10^{-2}$. The data is insensitive to such 
small deviation and therefore cannot probe the interplay between the dip and 
subsequent rise in the matter power (Fig.~\ref{fig:Ps-rel-diffs}). 
Fig.~\ref{fig:rel-diff-contours} implies that significant duration of USR 
$\Delta N \geq 0.5$, can occur only for $N_1 > 31$, 
or equivalently $\bar N_1 < 41$.
This means that substantial rise in power can occur only at $k \geq 10^2\,\mpcinv$.

We can understand our results using Eqs.~(\ref{eq:Pb3D}), (\ref{eq:pk1db}), and~(\ref{eq:jeans}). 
In the usual $\Lambda$CDM model based on PL power spectra from slow roll
inflationary models, the three-dimensional 
matter power spectrum scales as $k^{-3} (\log(k))^2$ at small scales (see e.g. \cite{dodelson}). For $k \lesssim k_{\rm J}$, the baryonic
1D power spectrum scales as $k^{-1}(\log(k))^2$ and for $k > k_{\rm J}$, $P^{{(1)}}_b (k, z) \propto k^{-5} (\log(k))^2$.  In the USR models, the increase in the primordial power spectrum can scale as $k^4$ (e.g. Fig.~\ref{fig:Ps-USR}).  
This means the main contribution to the enhanced power comes from scales $k \lesssim k_{\rm J} \simeq 10\,\mpcinv$ as the 1D baryonic power spectrum scales as $k (\log(k))^2$ for $k < k_{\rm J}$. However, for $k > k_{\rm J}$ the decrement in 1D power spectrum is shallow ($P^{{(1)}}_b (k, z) \propto k^{-1} (\log(k))^2$)  and we expect non-negligible 
contribution from smaller scales. Our analysis suggests scales up to 10 times smaller
than $k = k_{\rm J}$ can be constrained in USR models as the 1D power spectrum falls 
by only a factor of two in this range:
$P^{{(1)}}_b (100, z)/P^{{(1)}}_b (10, z) \simeq 0.5$\,.

\begin{figure}[!t]
\centering
\includegraphics[scale=0.5]{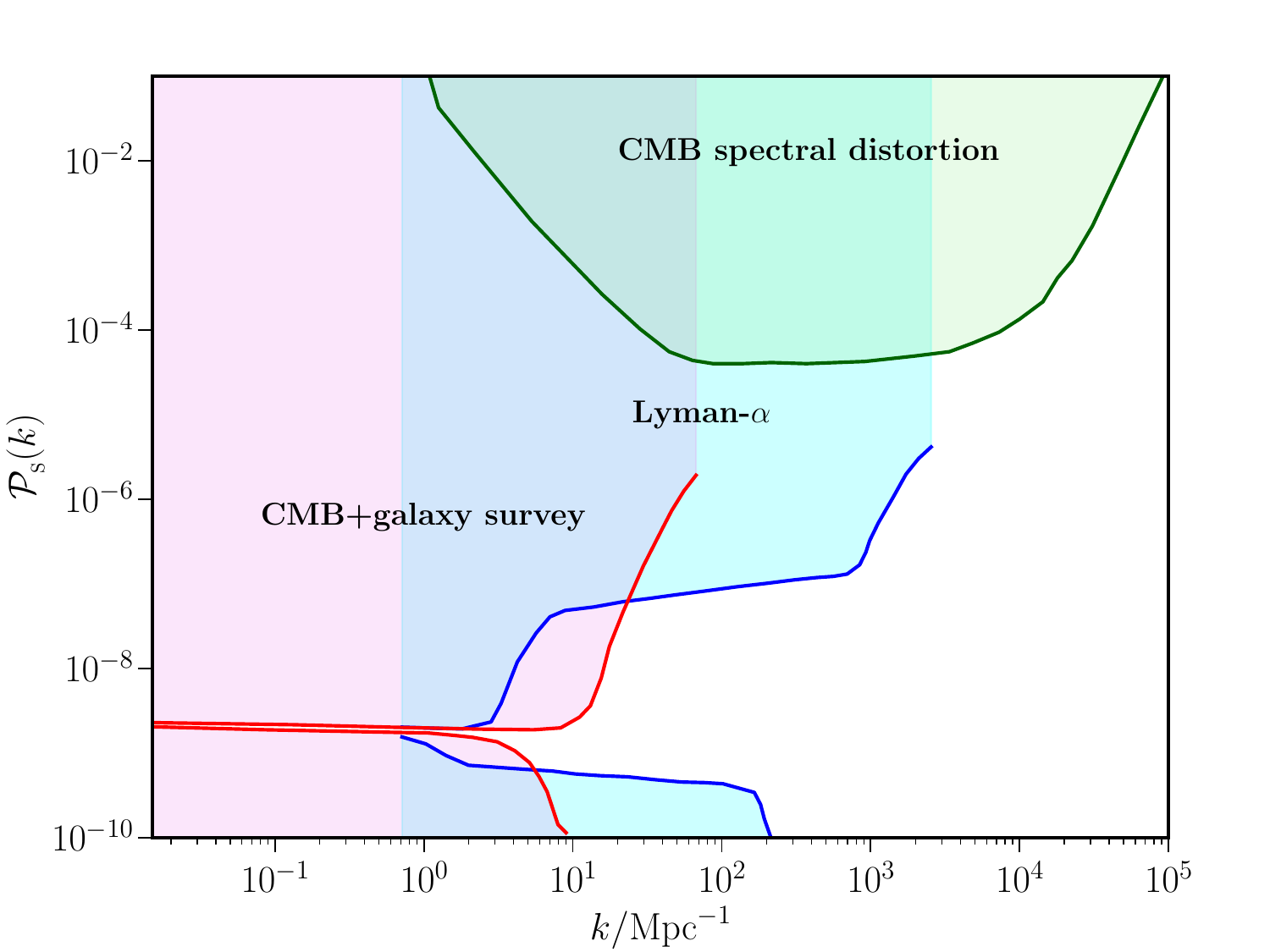}
\vskip -0.1in
\caption{The constraints shown in Fig.~\ref{fig:Ps-rel-diffs} are displayed in 
$\ps(k)\hbox{--}k$ plane. The red and blue lines correspond to CMB+galaxy 
surveys and Lyman-$\alpha$ data, respectively.
The region bounded by these lines, e.g. narrow band around $\ps(k) \simeq 2 \times 10^{-9}$, is allowed. 
The red lines demarcating the allowed region correspond to the envelopes of power
spectra presented in the left panel of Fig.~\ref{fig:Ps-rel-diffs}, which in
turn is obtained from the 2-$\sigma$ contour in the $N_1$-$\Delta N$ plane in 
Fig.~\ref{fig:USR_5_params_CMB+gal}. The blue lines bounding the allowed region are 
the envelopes of power spectra
presented in the right panel of Fig.~\ref{fig:Ps-rel-diffs}. We have taken the
set of spectra which lead to an average relative difference 
$\Delta ^{\rm 1D}_{\rm m}(z_\ast)=5$.
The three bands, from left to right, correspond to the region of 
spectrum ruled out by CMB+galaxy surveys data (lighter shade of red), both  
CMB+galaxy surveys and Lyman-$\alpha$ (in light blue), and just by Lyman-$\alpha$ 
data (shaded in cyan). 
We also display the constraints from CMB spectral distortion (green lines bounding  
the region in lighter shade of green)~\cite{Chluba:2019kpb}.}
\label{fig:Ps-constraints}
\end{figure}


\section{Conclusion}\label{sec:conc}
Recent LIGO and PTA  discoveries have opened the possibility of enhanced scalar power 
at small scales. This enhancement can be achieved by a brief USR phase during the inflationary era. 
In this paper, we aim to detect/constrain the dynamics of this phase of inflation using cosmological
data. In particular, we use CMB data owing to its unprecedented precision and Lyman-$\alpha$ data as it is  sensitive  to small scales. 
We parameterize our models to determine  the onset of the USR phase during inflation  and the deviation of the matter power as compared to the power-law model.  

From CMB and other data at comparable scales, we find that the earliest onset of USR 
can only be $\bar N_1 \lesssim 45$ 
(Fig.~\ref{fig:USR_5_params_CMB+gal}).
With Lyman-$\alpha$ dataset, the onset gets pushed further to 
$\bar N_1 \lesssim 41$ (Fig.~\ref{fig:rel-diff-contours}). 
The duration $\Delta N$ is largely unconstrained from CMB and other datasets. 
But Lyman-$\alpha$ strongly constrains the duration to be $\Delta N \lesssim 0.4$ 
along with  $\bar N_1 < 41$. These constraints limit the possibility of having any 
appreciable epoch of USR only to the last $41$ e-folds of inflation.
This implies a constraint on the amplitude of $\ps(k) < 10^{-7}$ over
$0.1 < k/\mpcinv < 10^2$. Such a constraint is a considerable improvement
of about two orders of magnitude on the bound on $\ps(k)$ compared to the limit
from {\tt FIRAS} over these scales~\cite{Chluba:2019kpb}.

In Fig.~\ref{fig:Ps-rel-diffs}, we plot the range of power spectra that are allowed by 
different data sets. In Fig.~\ref{fig:Ps-constraints}, we present our constraints
in terms of the primordial power spectrum and wavenumber, which is more readily connected to
cosmological observables and can be compared to other constraints such as spectral distortion. 
This helps us see the improvement in the constraint on the amplitude of power
by Lyman-$\alpha$ data  over spectral distortion for $k \lesssim 10^2\,\mpcinv$.
We should add that the exact shape of the bounds, especially the blue lines corresponding 
to Lyman-$\alpha$ constraints in Fig.~\ref{fig:Ps-constraints}, are dictated to some 
extent by the shape of the spectrum in the USR model of interest
(e.g.~\cite{2011MNRAS.413.1717B,Bringmann:2011ut,Kohri:2014lza,Nakama:2017qac} for similar 
constraints but for different models of dark matter and specific parameterized forms of 
the primordial power spectra.).

In light of our constraint on the onset of USR, we should remark about the 
implications of USR  phase for other cosmological datasets and the possibility of
extending these constraints using future observations.
\begin{enumerate}
\item 
As mentioned earlier, the dip in the tree-level power spectrum prior to rise is a unique signature of USR models. In an earlier work, 
it was shown that this dip could be gleaned in the spectrum of 21-cm signals from 
the Dark Ages~\cite{Balaji:2022zur}. 
Due to the constraint that we have obtained on $\bar N_1$ and $\Delta N$, if the 
dip were to occur at $k \simeq 10\,\mpcinv$, it shall be shallower, with the 
amplitude at the lowest point within an order of magnitude of the nearly 
scale-invariant behavior. Else it may occur deeper but only at $k \geq 10^2\,\mpcinv$.
Besides, the loop-level contributions that may alter the shape of the dip may also
exhibit their impact at such smaller scales~\cite{Franciolini:2023lgy}.

\item
To constrain the onset of USR at even smaller scales, one needs to explore 
 possible future constraints due to spectral distortions over the scales
of $10^4\,\mpcinv \lesssim k \lesssim 10^6\,\mpcinv$. We have already accounted for 
the constraint due to {\tt FIRAS} on $\ps(k) \lesssim 10^{-5}$ by restricting the 
range of duration of USR to be $\Delta N < 1$.
But an upcoming mission like {\tt PIXIE} with sensitivity as strong as 
$\ps(k) \sim 10^{-8}$ can strengthen our constraints on $\bar N_1$ and 
$\Delta N$~\cite{2011JCAP...07..025K}. 
Conversely the detection of spectral distortion of such levels, could be
a confirmation of a brief USR phase.

\item
As noted in the foregoing, there have been attempts to explain the stochastic
GWs detected by {\tt PTA} using USR models~\cite{Firouzjahi:2023lzg,Das:2023nmm,Choudhury:2024one}. 
A scalar power spectrum that is required to induce such a secondary GW spectrum 
should have its steep rise in power between $10^5\hbox{--}10^7\,\mpcinv$. This in turn 
corresponds to a phase of USR beginning at about $\bar N_1 \leq 32$\,. Such a value
is still in the viable range of $\bar N_1$ as per the bounds we have obtained.
If the signal by {\tt PTA} is indeed due to USR, then it provides a crucial insight 
into the onset and duration of USR phase during inflation. However, we should caution 
that such a steep enhancement in power is beget along with strong scalar 
non-Gaussianity, captured in a non-trivial behavior in the associated non-Gaussianity 
parameter $\fnl$~\cite{Ragavendra:2023ret}.
One needs to be careful in accounting for the non-Gaussian contributions to 
$\Omega_{\rm GW}$ while comparing USR models against GW datasets~\cite{Cai:2018dig,Unal:2018yaa,Atal:2021jyo,Adshead:2021hnm,Zhang:2021vak,Ragavendra:2021qdu,Garcia-Saenz:2022tzu,Chen:2022dah,Li:2023qua,Franciolini:2023pbf,Liu:2023ymk,Yuan:2023ofl,Li:2023xtl,Choudhury:2023fwk,Chang:2023aba,Inui:2023qsd,Chang:2023vjk,Papanikolaou:2024kjb}.

\item
The discovery of many  high-redshift galaxies with unusually high stellar mass by  JWST (e.g.~\cite{2023MNRAS.518.6011D,2023Natur.616..266L}) could be pointing at cosmological  models with more power at small scales as compared to the $\Lambda$CDM model. However, it is also likely that the observed behaviour is owing to much higher star formation efficiency at high redshifts with significant contribution from population III stars (e.g.~\cite{2023MNRAS.518.6011D} and references therein). This issue is still being debated and remains unsettled.
\end{enumerate}

In this paper, we consider  the tree-level contribution to the primordial power 
spectrum. For USR models, there is ongoing discussion in the literature about the 
possibility of higher-order (loop-level) contribution dominating over
the tree-level primordial power spectrum (e.g.~\cite{Syu:2019uwx,Kristiano:2022maq,Inomata:2022yte,Choudhury:2023vuj,Firouzjahi:2023bkt,Cheng:2023ikq,Choudhury:2023rks,Maity:2023qzw}
and \cite{Riotto:2023hoz,Firouzjahi:2023ahg,Fumagalli:2023hpa,Cheng:2021lif} 
for counter claims).
Within the framework of this ongoing debate, such an effect is possible only in 
the following cases:  (i) the transitions between SR and 
USR are near-instantaneous and (ii) the enhancement in power is such that 
$\ps(k) \sim 10^{-2}$, as required in the context of production of PBHs.
This is not the case in our model as the transitions are not instantaneous, but
smooth as is the case in any realistic model of USR driven by a smooth potential.
Also, the range of $\Delta N$ we have worked with is restricted such that the
maximum amplitude of $\ps(k) \leq 10^{-5}$ as we do not violate the bound due to
spectral distortion. Hence, the argument of loop-level spectrum dominating over 
tree-level spectrum does not occur in our analysis.


\section*{Acknowledgements}
The authors thank Guillem Dom\`{e}nech and Ujjwal Kumar Upadhyay for useful comments and discussion.
The authors acknowledge the usage of high-performance computing cluster at Raman 
Research Institute for various numerical computations.
HVR thanks Raman Research Institute for support through postdoctoral fellowship.

\bibliographystyle{JHEP}
\bibliography{usr}

\providecommand{\href}[2]{#2}\begingroup\raggedright\begin{thebibliography}{100}

\bibitem{Ozsoy:2023ryl}
O.~\"Ozsoy and G.~Tasinato, {\it {Inflation and Primordial Black Holes}},  {\em
  Universe} {\bf 9} (2023), no.~5 203,
  [\href{http://arxiv.org/abs/2301.03600}{{\tt arXiv:2301.03600}}].

\bibitem{Domenech:2021ztg}
G.~Dom\`enech, {\it {Scalar Induced Gravitational Waves Review}},  {\em
  Universe} {\bf 7} (2021), no.~11 398,
  [\href{http://arxiv.org/abs/2109.01398}{{\tt arXiv:2109.01398}}].

\bibitem{LISACosmologyWorkingGroup:2023njw}
{\bf LISA Cosmology Working Group} Collaboration, E.~Bagui et~al., {\it
  {Primordial black holes and their gravitational-wave signatures}},
  \href{http://arxiv.org/abs/2310.19857}{{\tt arXiv:2310.19857}}.

\bibitem{LIGOScientific:2018mvr}
{\bf LIGO Scientific, Virgo} Collaboration, B.~P. Abbott et~al., {\it {GWTC-1:
  A Gravitational-Wave Transient Catalog of Compact Binary Mergers Observed by
  LIGO and Virgo during the First and Second Observing Runs}},  {\em Phys. Rev.
  X} {\bf 9} (2019), no.~3 031040, [\href{http://arxiv.org/abs/1811.12907}{{\tt
  arXiv:1811.12907}}].

\bibitem{LIGOScientific:2021usb}
{\bf LIGO Scientific, VIRGO} Collaboration, R.~Abbott et~al., {\it {GWTC-2.1:
  Deep Extended Catalog of Compact Binary Coalescences Observed by LIGO and
  Virgo During the First Half of the Third Observing Run}},
  \href{http://arxiv.org/abs/2108.01045}{{\tt arXiv:2108.01045}}.

\bibitem{LIGOScientific:2021djp}
{\bf LIGO Scientific, VIRGO, KAGRA} Collaboration, R.~Abbott et~al., {\it
  {GWTC-3: Compact Binary Coalescences Observed by LIGO and Virgo During the
  Second Part of the Third Observing Run}},
  \href{http://arxiv.org/abs/2111.03606}{{\tt arXiv:2111.03606}}.

\bibitem{NANOGrav:2023gor}
{\bf NANOGrav} Collaboration, G.~Agazie et~al., {\it {The NANOGrav 15 yr Data
  Set: Evidence for a Gravitational-wave Background}},  {\em Astrophys. J.
  Lett.} {\bf 951} (2023), no.~1 L8,
  [\href{http://arxiv.org/abs/2306.16213}{{\tt arXiv:2306.16213}}].

\bibitem{EPTA:2023fyk}
{\bf EPTA, InPTA:} Collaboration, J.~Antoniadis et~al., {\it {The second data
  release from the European Pulsar Timing Array - III. Search for gravitational
  wave signals}},  {\em Astron. Astrophys.} {\bf 678} (2023) A50,
  [\href{http://arxiv.org/abs/2306.16214}{{\tt arXiv:2306.16214}}].

\bibitem{Bird:2016dcv}
S.~Bird, I.~Cholis, J.~B. Muñoz, Y.~Ali-Haïmoud, M.~Kamionkowski, E.~D.
  Kovetz, A.~Raccanelli, and A.~G. Riess, {\it {Did LIGO detect dark matter?}},
   {\em Phys. Rev. Lett.} {\bf 116} (2016), no.~20 201301,
  [\href{http://arxiv.org/abs/1603.00464}{{\tt arXiv:1603.00464}}].

\bibitem{Sasaki:2016jop}
M.~Sasaki, T.~Suyama, T.~Tanaka, and S.~Yokoyama, {\it {Primordial Black Hole
  Scenario for the Gravitational-Wave Event GW150914}},  {\em Phys. Rev. Lett.}
  {\bf 117} (2016), no.~6 061101, [\href{http://arxiv.org/abs/1603.08338}{{\tt
  arXiv:1603.08338}}]. [erratum: Phys. Rev. Lett.121,no.5,059901(2018)].

\bibitem{DeLuca:2020qqa}
V.~De~Luca, G.~Franciolini, P.~Pani, and A.~Riotto, {\it {Primordial Black
  Holes Confront LIGO/Virgo data: Current situation}},  {\em JCAP} {\bf 06}
  (2020) 044, [\href{http://arxiv.org/abs/2005.05641}{{\tt arXiv:2005.05641}}].

\bibitem{Jedamzik:2020ypm}
K.~Jedamzik, {\it {Primordial Black Hole Dark Matter and the LIGO/Virgo
  observations}},  {\em JCAP} {\bf 09} (2020) 022,
  [\href{http://arxiv.org/abs/2006.11172}{{\tt arXiv:2006.11172}}].

\bibitem{Jedamzik:2020omx}
K.~Jedamzik, {\it {Consistency of Primordial Black Hole Dark Matter with
  LIGO/Virgo Merger Rates}},  {\em Phys. Rev. Lett.} {\bf 126} (2021), no.~5
  051302, [\href{http://arxiv.org/abs/2007.03565}{{\tt arXiv:2007.03565}}].

\bibitem{Chen:2024dxh}
Z.-C. Chen and A.~Hall, {\it {Confronting primordial black holes with
  LIGO-Virgo-KAGRA and the Einstein Telescope}},
  \href{http://arxiv.org/abs/2402.03934}{{\tt arXiv:2402.03934}}.

\bibitem{NANOGrav:2023hvm}
{\bf NANOGrav} Collaboration, A.~Afzal et~al., {\it {The NANOGrav 15 yr Data
  Set: Search for Signals from New Physics}},  {\em Astrophys. J. Lett.} {\bf
  951} (2023), no.~1 L11, [\href{http://arxiv.org/abs/2306.16219}{{\tt
  arXiv:2306.16219}}].

\bibitem{Cai:2023dls}
Y.-F. Cai, X.-C. He, X.-H. Ma, S.-F. Yan, and G.-W. Yuan, {\it {Limits on
  scalar-induced gravitational waves from the stochastic background by pulsar
  timing array observations}},  {\em Sci. Bull.} {\bf 68} (2023) 2929--2935,
  [\href{http://arxiv.org/abs/2306.17822}{{\tt arXiv:2306.17822}}].

\bibitem{Vagnozzi:2023lwo}
S.~Vagnozzi, {\it {Inflationary interpretation of the stochastic gravitational
  wave background signal detected by pulsar timing array experiments}},  {\em
  JHEAp} {\bf 39} (2023) 81--98, [\href{http://arxiv.org/abs/2306.16912}{{\tt
  arXiv:2306.16912}}].

\bibitem{Choudhury:2023kam}
S.~Choudhury, {\it {Single field inflation in the light of Pulsar Timing Array
  Data: Quintessential interpretation of blue tilted tensor spectrum through
  Non-Bunch Davies initial condition}},  {\em Eur. Phys. J. C} {\bf 84} (2024)
  278, [\href{http://arxiv.org/abs/2307.03249}{{\tt arXiv:2307.03249}}].

\bibitem{Firouzjahi:2023lzg}
H.~Firouzjahi and A.~Talebian, {\it {Induced gravitational waves from ultra
  slow-roll inflation and pulsar timing arrays observations}},  {\em JCAP} {\bf
  10} (2023) 032, [\href{http://arxiv.org/abs/2307.03164}{{\tt
  arXiv:2307.03164}}].

\bibitem{Das:2023nmm}
B.~Das, N.~Jaman, and M.~Sami, {\it {Gravitational wave background from
  quintessential inflation and NANOGrav data}},  {\em Phys. Rev. D} {\bf 108}
  (2023), no.~10 103510, [\href{http://arxiv.org/abs/2307.12913}{{\tt
  arXiv:2307.12913}}].

\bibitem{Liu:2023pau}
L.~Liu, Z.-C. Chen, and Q.-G. Huang, {\it {Probing the equation of state of the
  early Universe with pulsar timing arrays}},  {\em JCAP} {\bf 11} (2023) 071,
  [\href{http://arxiv.org/abs/2307.14911}{{\tt arXiv:2307.14911}}].

\bibitem{Ellis:2023oxs}
J.~Ellis, M.~Fairbairn, G.~Franciolini, G.~H\"utsi, A.~Iovino, M.~Lewicki,
  M.~Raidal, J.~Urrutia, V.~Vaskonen, and H.~Veerm\"ae, {\it {What is the
  source of the PTA GW signal?}},  {\em Phys. Rev. D} {\bf 109} (2024), no.~2
  023522, [\href{http://arxiv.org/abs/2308.08546}{{\tt arXiv:2308.08546}}].

\bibitem{Gangopadhyay:2023qjr}
M.~R. Gangopadhyay, V.~V. Godithi, K.~Ichiki, R.~Inui, T.~Kajino,
  A.~Manusankar, G.~J. Mathews, and Yogesh, {\it {Is the NANOGrav detection
  evidence of resonant particle creation during inflation?}},
  \href{http://arxiv.org/abs/2309.03101}{{\tt arXiv:2309.03101}}.

\bibitem{Fei:2023iel}
Q.~Fei, {\it {Constraints on the primordial curvature power spectrum by pulsar
  timing array data: a polynomial parameterization approach}},  {\em Commun.
  Theor. Phys.} {\bf 76} (2024), no.~1 015404,
  [\href{http://arxiv.org/abs/2310.17199}{{\tt arXiv:2310.17199}}].

\bibitem{Choudhury:2024one}
S.~Choudhury, A.~Karde, S.~Panda, and M.~Sami, {\it {Realisation of the
  ultra-slow roll phase in Galileon inflation and PBH overproduction}},
  \href{http://arxiv.org/abs/2401.10925}{{\tt arXiv:2401.10925}}.

\bibitem{Chen:2024twp}
Z.-C. Chen and L.~Liu, {\it {Can we distinguish the adiabatic fluctuations and
  isocurvature fluctuations with pulsar timing arrays?}},
  \href{http://arxiv.org/abs/2402.16781}{{\tt arXiv:2402.16781}}.

\bibitem{Starobinsky:1992ts}
A.~A. Starobinsky, {\it {Spectrum of adiabatic perturbations in the universe
  when there are singularities in the inflation potential}},  {\em JETP Lett.}
  {\bf 55} (1992) 489--494.

\bibitem{Hazra:2010ve}
D.~K. Hazra, M.~Aich, R.~K. Jain, L.~Sriramkumar, and T.~Souradeep, {\it
  {Primordial features due to a step in the inflaton potential}},  {\em JCAP}
  {\bf 10} (2010) 008, [\href{http://arxiv.org/abs/1005.2175}{{\tt
  arXiv:1005.2175}}].

\bibitem{Martin:2014kja}
J.~Martin, L.~Sriramkumar, and D.~K. Hazra, {\it {Sharp inflaton potentials and
  bi-spectra: Effects of smoothening the discontinuity}},  {\em JCAP} {\bf 09}
  (2014) 039, [\href{http://arxiv.org/abs/1404.6093}{{\tt arXiv:1404.6093}}].

\bibitem{Garcia-Bellido:2017mdw}
J.~Garcia-Bellido and E.~Ruiz~Morales, {\it {Primordial black holes from single
  field models of inflation}},  {\em Phys. Dark Univ.} {\bf 18} (2017) 47--54,
  [\href{http://arxiv.org/abs/1702.03901}{{\tt arXiv:1702.03901}}].

\bibitem{Germani:2017bcs}
C.~Germani and T.~Prokopec, {\it {On primordial black holes from an inflection
  point}},  {\em Phys. Dark Univ.} {\bf 18} (2017) 6--10,
  [\href{http://arxiv.org/abs/1706.04226}{{\tt arXiv:1706.04226}}].

\bibitem{Ballesteros:2017fsr}
G.~Ballesteros and M.~Taoso, {\it {Primordial black hole dark matter from
  single field inflation}},  {\em Phys. Rev. D} {\bf 97} (2018), no.~2 023501,
  [\href{http://arxiv.org/abs/1709.05565}{{\tt arXiv:1709.05565}}].

\bibitem{Ezquiaga:2017fvi}
J.~M. Ezquiaga, J.~Garcia-Bellido, and E.~Ruiz~Morales, {\it {Primordial Black
  Hole production in Critical Higgs Inflation}},  {\em Phys. Lett.} {\bf B776}
  (2018) 345--349, [\href{http://arxiv.org/abs/1705.04861}{{\tt
  arXiv:1705.04861}}].

\bibitem{Bezrukov:2017dyv}
F.~Bezrukov, M.~Pauly, and J.~Rubio, {\it {On the robustness of the primordial
  power spectrum in renormalized Higgs inflation}},  {\em JCAP} {\bf 02} (2018)
  040, [\href{http://arxiv.org/abs/1706.05007}{{\tt arXiv:1706.05007}}].

\bibitem{Drees:2019xpp}
M.~Drees and Y.~Xu, {\it {Overshooting, Critical Higgs Inflation and Second
  Order Gravitational Wave Signatures}},  {\em Eur. Phys. J. C} {\bf 81}
  (2021), no.~2 182, [\href{http://arxiv.org/abs/1905.13581}{{\tt
  arXiv:1905.13581}}].

\bibitem{Atal:2019cdz}
V.~Atal, J.~Garriga, and A.~Marcos-Caballero, {\it {Primordial black hole
  formation with non-Gaussian curvature perturbations}},  {\em JCAP} {\bf 09}
  (2019) 073, [\href{http://arxiv.org/abs/1905.13202}{{\tt arXiv:1905.13202}}].

\bibitem{Mishra:2019pzq}
S.~S. Mishra and V.~Sahni, {\it {Primordial Black Holes from a tiny bump/dip in
  the Inflaton potential}},  {\em JCAP} {\bf 04} (2020) 007,
  [\href{http://arxiv.org/abs/1911.00057}{{\tt arXiv:1911.00057}}].

\bibitem{Ballesteros:2020qam}
G.~Ballesteros, J.~Rey, M.~Taoso, and A.~Urbano, {\it {Primordial black holes
  as dark matter and gravitational waves from single-field polynomial
  inflation}},  \href{http://arxiv.org/abs/2001.08220}{{\tt arXiv:2001.08220}}.

\bibitem{Kefala:2020xsx}
K.~Kefala, G.~P. Kodaxis, I.~D. Stamou, and N.~Tetradis, {\it {Features of the
  inflaton potential and the power spectrum of cosmological perturbations}},
  {\em Phys. Rev. D} {\bf 104} (2021), no.~2 023506,
  [\href{http://arxiv.org/abs/2010.12483}{{\tt arXiv:2010.12483}}].

\bibitem{Braglia:2020eai}
M.~Braglia, D.~K. Hazra, F.~Finelli, G.~F. Smoot, L.~Sriramkumar, and A.~A.
  Starobinsky, {\it {Generating PBHs and small-scale GWs in two-field models of
  inflation}},  {\em JCAP} {\bf 08} (2020) 001,
  [\href{http://arxiv.org/abs/2005.02895}{{\tt arXiv:2005.02895}}].

\bibitem{ZhengRuiFeng:2021zoz}
R.~Zheng and T.~Q. Shi, Jiaming~and, {\it {On primordial black holes and
  secondary gravitational waves generated from inflation with solo/multi-bumpy
  potential *}},  {\em Chin. Phys. C} {\bf 46} (2022), no.~4 045103,
  [\href{http://arxiv.org/abs/2106.04303}{{\tt arXiv:2106.04303}}].

\bibitem{Braglia:2022phb}
M.~Braglia, A.~Linde, R.~Kallosh, and F.~Finelli, {\it {Hybrid
  \ensuremath{\alpha}-attractors, primordial black holes and gravitational wave
  backgrounds}},  {\em JCAP} {\bf 04} (2023) 033,
  [\href{http://arxiv.org/abs/2211.14262}{{\tt arXiv:2211.14262}}].

\bibitem{Zhao:2023zbg}
H.-R. Zhao, Y.-C. Liu, J.-X. Zhao, and N.~Li, {\it {The evolution of the
  primordial curvature perturbation in the ultraslow-roll inflation}},  {\em
  Eur. Phys. J. C} {\bf 83} (2023), no.~9 783.

\bibitem{Choudhury:2023hfm}
S.~Choudhury, A.~Karde, S.~Panda, and M.~Sami, {\it {Scalar induced gravity
  waves from ultra slow-roll Galileon inflation}},
  \href{http://arxiv.org/abs/2308.09273}{{\tt arXiv:2308.09273}}.

\bibitem{Ragavendra:2023ret}
H.~V. Ragavendra and L.~Sriramkumar, {\it {Observational Imprints of Enhanced
  Scalar Power on Small Scales in Ultra Slow Roll Inflation and Associated
  Non-Gaussianities}},  {\em Galaxies} {\bf 11} (2023), no.~1 34,
  [\href{http://arxiv.org/abs/2301.08887}{{\tt arXiv:2301.08887}}].

\bibitem{Byrnes:2018txb}
C.~T. Byrnes, P.~S. Cole, and S.~P. Patil, {\it {Steepest growth of the power
  spectrum and primordial black holes}},  {\em JCAP} {\bf 06} (2019) 028,
  [\href{http://arxiv.org/abs/1811.11158}{{\tt arXiv:1811.11158}}].

\bibitem{Motohashi:2019rhu}
H.~Motohashi, S.~Mukohyama, and M.~Oliosi, {\it {Constant Roll and Primordial
  Black Holes}},  {\em JCAP} {\bf 03} (2020) 002,
  [\href{http://arxiv.org/abs/1910.13235}{{\tt arXiv:1910.13235}}].

\bibitem{Tasinato:2020vdk}
G.~Tasinato, {\it {An analytic approach to non-slow-roll inflation}},  {\em
  Phys. Rev. D} {\bf 103} (2021), no.~2 023535,
  [\href{http://arxiv.org/abs/2012.02518}{{\tt arXiv:2012.02518}}].

\bibitem{Ragavendra:2020sop}
H.~V. Ragavendra, P.~Saha, L.~Sriramkumar, and J.~Silk, {\it {Primordial black
  holes and secondary gravitational waves from ultraslow roll and punctuated
  inflation}},  {\em Phys. Rev. D} {\bf 103} (2021), no.~8 083510,
  [\href{http://arxiv.org/abs/2008.12202}{{\tt arXiv:2008.12202}}].

\bibitem{Franciolini:2022pav}
G.~Franciolini and A.~Urbano, {\it {Primordial black hole dark matter from
  inflation: The reverse engineering approach}},  {\em Phys. Rev. D} {\bf 106}
  (2022), no.~12 123519, [\href{http://arxiv.org/abs/2207.10056}{{\tt
  arXiv:2207.10056}}].

\bibitem{Domenech:2023dxx}
G.~Dom\`enech, G.~Vargas, and T.~Vargas, {\it {An exact model for
  enhancing/suppressing primordial fluctuations}},  {\em JCAP} {\bf 03} (2024)
  002, [\href{http://arxiv.org/abs/2309.05750}{{\tt arXiv:2309.05750}}].

\bibitem{Ozsoy:2019lyy}
O.~\"Ozsoy and G.~Tasinato, {\it {On the slope of the curvature power spectrum
  in non-attractor inflation}},  {\em JCAP} {\bf 04} (2020) 048,
  [\href{http://arxiv.org/abs/1912.01061}{{\tt arXiv:1912.01061}}].

\bibitem{Balaji:2022zur}
S.~Balaji, H.~V. Ragavendra, S.~K. Sethi, J.~Silk, and L.~Sriramkumar, {\it
  {Observing Nulling of Primordial Correlations via the 21-cm Signal}},  {\em
  Phys. Rev. Lett.} {\bf 129} (2022), no.~26 261301,
  [\href{http://arxiv.org/abs/2206.06386}{{\tt arXiv:2206.06386}}].

\bibitem{Franciolini:2023lgy}
G.~Franciolini, A.~Iovino, Junior., M.~Taoso, and A.~Urbano, {\it {One loop to
  rule them all: Perturbativity in the presence of ultra slow-roll dynamics}},
  \href{http://arxiv.org/abs/2305.03491}{{\tt arXiv:2305.03491}}.

\bibitem{Ng:2021hll}
K.-W. Ng and Y.-P. Wu, {\it {Constant-rate inflation: primordial black holes
  from conformal weight transitions}},  {\em JHEP} {\bf 11} (2021) 076,
  [\href{http://arxiv.org/abs/2102.05620}{{\tt arXiv:2102.05620}}].

\bibitem{Cole:2022xqc}
P.~S. Cole, A.~D. Gow, C.~T. Byrnes, and S.~P. Patil, {\it {Steepest growth
  re-examined: repercussions for primordial black hole formation}},
  \href{http://arxiv.org/abs/2204.07573}{{\tt arXiv:2204.07573}}.

\bibitem{Bhaumik:2019tvl}
N.~Bhaumik and R.~K. Jain, {\it {Primordial black holes dark matter from
  inflection point models of inflation and the effects of reheating}},
  \href{http://arxiv.org/abs/1907.04125}{{\tt arXiv:1907.04125}}.
  [JCAP2001,037(2020)].

\bibitem{Iacconi:2021ltm}
L.~Iacconi, H.~Assadullahi, M.~Fasiello, and D.~Wands, {\it {Revisiting
  small-scale fluctuations in \ensuremath{\alpha}-attractor models of
  inflation}},  {\em JCAP} {\bf 06} (2022), no.~06 007,
  [\href{http://arxiv.org/abs/2112.05092}{{\tt arXiv:2112.05092}}].

\bibitem{Cicoli:2022sih}
M.~Cicoli, F.~G. Pedro, and N.~Pedron, {\it {Secondary GWs and PBHs in string
  inflation: formation and detectability}},  {\em JCAP} {\bf 08} (2022), no.~08
  030, [\href{http://arxiv.org/abs/2203.00021}{{\tt arXiv:2203.00021}}].

\bibitem{Karam:2022nym}
A.~Karam, N.~Koivunen, E.~Tomberg, V.~Vaskonen, and H.~Veerm\"ae, {\it {Anatomy
  of single-field inflationary models for primordial black holes}},  {\em JCAP}
  {\bf 03} (2023) 013, [\href{http://arxiv.org/abs/2205.13540}{{\tt
  arXiv:2205.13540}}].

\bibitem{Qin:2023lgo}
W.~Qin, S.~R. Geller, S.~Balaji, E.~McDonough, and D.~I. Kaiser, {\it {Planck
  constraints and gravitational wave forecasts for primordial black hole dark
  matter seeded by multifield inflation}},  {\em Phys. Rev. D} {\bf 108}
  (2023), no.~4 043508, [\href{http://arxiv.org/abs/2303.02168}{{\tt
  arXiv:2303.02168}}].

\bibitem{Tagliazucchi:2023dai}
M.~Tagliazucchi, M.~Braglia, F.~Finelli, and M.~Pieroni, {\it {The quest of CMB
  spectral distortions to probe the scalar-induced gravitational wave
  background interpretation in PTA data}},
  \href{http://arxiv.org/abs/2310.08527}{{\tt arXiv:2310.08527}}.

\bibitem{Kohri:2007qn}
K.~Kohri, D.~H. Lyth, and A.~Melchiorri, {\it {Black hole formation and
  slow-roll inflation}},  {\em JCAP} {\bf 0804} (2008) 038,
  [\href{http://arxiv.org/abs/0711.5006}{{\tt arXiv:0711.5006}}].

\bibitem{Mather:1993ij}
J.~C. Mather et~al., {\it {Measurement of the Cosmic Microwave Background
  spectrum by the COBE FIRAS instrument}},  {\em Astrophys. J.} {\bf 420}
  (1994) 439--444.

\bibitem{Fixsen:1996nj}
D.~J. Fixsen, E.~S. Cheng, J.~M. Gales, J.~C. Mather, R.~A. Shafer, and E.~L.
  Wright, {\it {The Cosmic Microwave Background spectrum from the full COBE
  FIRAS data set}},  {\em Astrophys. J.} {\bf 473} (1996) 576,
  [\href{http://arxiv.org/abs/astro-ph/9605054}{{\tt astro-ph/9605054}}].

\bibitem{faucher2008direct}
C.-A. Faucher-Giguere, J.~X. Prochaska, A.~Lidz, L.~Hernquist, and
  M.~Zaldarriaga, {\it A direct precision measurement of the intergalactic
  ly$\alpha$ opacity at 2 $\leq z \leq$ 4.2},  {\em The Astrophysical Journal}
  {\bf 681} (2008), no.~2 831.

\bibitem{Mukhanov:1990me}
V.~F. Mukhanov, H.~A. Feldman, and R.~H. Brandenberger, {\it {Theory of
  cosmological perturbations. Part 1. Classical perturbations. Part 2. Quantum
  theory of perturbations. Part 3. Extensions}},  {\em Phys. Rept.} {\bf 215}
  (1992) 203--333.

\bibitem{Martin:2004um}
J.~Martin, {\it {Inflationary cosmological perturbations of quantum-mechanical
  origin}},  {\em Lect. Notes Phys.} {\bf 669} (2005) 199--244,
  [\href{http://arxiv.org/abs/hep-th/0406011}{{\tt hep-th/0406011}}].

\bibitem{Kinney:2009vz}
W.~H. Kinney, {\it {TASI Lectures on Inflation}},
  \href{http://arxiv.org/abs/0902.1529}{{\tt arXiv:0902.1529}}.

\bibitem{Baumann:2009ds}
D.~Baumann, {\it {Inflation}},  in {\em {Physics of the large and the small,
  TASI 09, proceedings of the Theoretical Advanced Study Institute in
  Elementary Particle Physics, Boulder, Colorado, USA, 1-26 June 2009}},
  pp.~523--686, 2011.
\newblock \href{http://arxiv.org/abs/0907.5424}{{\tt arXiv:0907.5424}}.

\bibitem{Sriramkumar:2009kg}
L.~Sriramkumar, {\it {An introduction to inflation and cosmological
  perturbation theory}},  \href{http://arxiv.org/abs/0904.4584}{{\tt
  arXiv:0904.4584}}.

\bibitem{Carrilho:2019oqg}
P.~Carrilho, K.~A. Malik, and D.~J. Mulryne, {\it {Dissecting the growth of the
  power spectrum for primordial black holes}},  {\em Phys. Rev. D} {\bf 100}
  (2019), no.~10 103529, [\href{http://arxiv.org/abs/1907.05237}{{\tt
  arXiv:1907.05237}}].

\bibitem{Ozsoy:2018flq}
O.~Özsoy, S.~Parameswaran, G.~Tasinato, and I.~Zavala, {\it {Mechanisms for
  Primordial Black Hole Production in String Theory}},  {\em JCAP} {\bf 1807}
  (2018) 005, [\href{http://arxiv.org/abs/1803.07626}{{\tt arXiv:1803.07626}}].

\bibitem{Ozsoy:2021pws}
O.~\"Ozsoy and G.~Tasinato, {\it {Consistency conditions and primordial black
  holes in single field inflation}},  {\em Phys. Rev. D} {\bf 105} (2022),
  no.~2 023524, [\href{http://arxiv.org/abs/2111.02432}{{\tt
  arXiv:2111.02432}}].

\bibitem{Liu:2020oqe}
J.~Liu, Z.-K. Guo, and R.-G. Cai, {\it {Analytical approximation of the scalar
  spectrum in the ultraslow-roll inflationary models}},  {\em Phys. Rev. D}
  {\bf 101} (2020), no.~8 083535, [\href{http://arxiv.org/abs/2003.02075}{{\tt
  arXiv:2003.02075}}].

\bibitem{Jain:2008dw}
R.~K. Jain, P.~Chingangbam, J.-O. Gong, L.~Sriramkumar, and T.~Souradeep, {\it
  {Punctuated inflation and the low CMB multipoles}},  {\em JCAP} {\bf 01}
  (2009) 009, [\href{http://arxiv.org/abs/0809.3915}{{\tt arXiv:0809.3915}}].

\bibitem{Jain:2009pm}
R.~K. Jain, P.~Chingangbam, L.~Sriramkumar, and T.~Souradeep, {\it {The
  tensor-to-scalar ratio in punctuated inflation}},  {\em Phys. Rev. D} {\bf
  82} (2010) 023509, [\href{http://arxiv.org/abs/0904.2518}{{\tt
  arXiv:0904.2518}}].

\bibitem{Qureshi:2016pjy}
M.~H. Qureshi, A.~Iqbal, M.~A. Malik, and T.~Souradeep, {\it {Low-$\ell$ power
  suppression in punctuated inflation}},  {\em JCAP} {\bf 04} (2017) 013,
  [\href{http://arxiv.org/abs/1610.05776}{{\tt arXiv:1610.05776}}].

\bibitem{Ragavendra:2020old}
H.~V. Ragavendra, D.~Chowdhury, and L.~Sriramkumar, {\it {Suppression of scalar
  power on large scales and associated bispectra}},  {\em Phys. Rev. D} {\bf
  106} (2022), no.~4 043535, [\href{http://arxiv.org/abs/2003.01099}{{\tt
  arXiv:2003.01099}}].

\bibitem{Planck:2019nip}
{\bf Planck} Collaboration, N.~Aghanim et~al., {\it {Planck 2018 results. V.
  CMB power spectra and likelihoods}},  {\em Astron. Astrophys.} {\bf 641}
  (2020) A5, [\href{http://arxiv.org/abs/1907.12875}{{\tt arXiv:1907.12875}}].

\bibitem{BICEP:2021xfz}
{\bf BICEP, Keck} Collaboration, P.~A.~R. Ade et~al., {\it {Improved
  Constraints on Primordial Gravitational Waves using Planck, WMAP, and
  BICEP/Keck Observations through the 2018 Observing Season}},  {\em Phys. Rev.
  Lett.} {\bf 127} (2021), no.~15 151301,
  [\href{http://arxiv.org/abs/2110.00483}{{\tt arXiv:2110.00483}}].

\bibitem{BOSS:2016wmc}
{\bf BOSS} Collaboration, S.~Alam et~al., {\it {The clustering of galaxies in
  the completed SDSS-III Baryon Oscillation Spectroscopic Survey: cosmological
  analysis of the DR12 galaxy sample}},  {\em Mon. Not. Roy. Astron. Soc.} {\bf
  470} (2017), no.~3 2617--2652, [\href{http://arxiv.org/abs/1607.03155}{{\tt
  arXiv:1607.03155}}].

\bibitem{Drlica-Wagner_2018}
A.~Drlica-Wagner, I.~Sevilla-Noarbe, E.~S. Rykoff, R.~A. Gruendl, B.~Yanny,
  D.~L. Tucker, B.~Hoyle, A.~C. Rosell, G.~M. Bernstein, K.~Bechtol, M.~R.
  Becker, A.~Benoit-Lévy, E.~Bertin, M.~C. Kind, C.~Davis, J.~de~Vicente,
  H.~T. Diehl, D.~Gruen, W.~G. Hartley, B.~Leistedt, T.~S. Li, J.~L. Marshall,
  E.~Neilsen, M.~M. Rau, E.~Sheldon, J.~Smith, M.~A. Troxel, S.~Wyatt,
  Y.~Zhang, T.~M.~C. Abbott, F.~B. Abdalla, S.~Allam, M.~Banerji, D.~Brooks,
  E.~Buckley-Geer, D.~L. Burke, D.~Capozzi, J.~Carretero, C.~E. Cunha, C.~B.
  D’Andrea, L.~N. da~Costa, D.~L. DePoy, S.~Desai, J.~P. Dietrich, P.~Doel,
  A.~E. Evrard, A.~F. Neto, B.~Flaugher, P.~Fosalba, J.~Frieman,
  J.~García-Bellido, D.~W. Gerdes, T.~Giannantonio, J.~Gschwend, G.~Gutierrez,
  K.~Honscheid, D.~J. James, T.~Jeltema, K.~Kuehn, S.~Kuhlmann, N.~Kuropatkin,
  O.~Lahav, M.~Lima, H.~Lin, M.~A.~G. Maia, P.~Martini, R.~G. McMahon,
  P.~Melchior, F.~Menanteau, R.~Miquel, R.~C. Nichol, R.~L.~C. Ogando, A.~A.
  Plazas, A.~K. Romer, A.~Roodman, E.~Sanchez, V.~Scarpine, R.~Schindler,
  M.~Schubnell, M.~Smith, R.~C. Smith, M.~Soares-Santos, F.~Sobreira,
  E.~Suchyta, G.~Tarle, V.~Vikram, A.~R. Walker, R.~H. Wechsler, J.~Zuntz, and
  D.~Collaboration), {\it Dark energy survey year 1 results: The photometric
  data set for cosmology},  {\em The Astrophysical Journal Supplement Series}
  {\bf 235} (apr, 2018) 33.

\bibitem{Lewis:2002ah}
A.~Lewis and S.~Bridle, {\it {Cosmological parameters from CMB and other data:
  A Monte Carlo approach}},  {\em Phys. Rev. D} {\bf 66} (2002) 103511,
  [\href{http://arxiv.org/abs/astro-ph/0205436}{{\tt astro-ph/0205436}}].

\bibitem{Lewis:1999bs}
A.~Lewis, A.~Challinor, and A.~Lasenby, {\it {Efficient computation of CMB
  anisotropies in closed FRW models}},  {\em Astrophys. J.} {\bf 538} (2000)
  473--476, [\href{http://arxiv.org/abs/astro-ph/9911177}{{\tt
  astro-ph/9911177}}].

\bibitem{Lewis:2019xzd}
A.~Lewis, {\it {GetDist: a Python package for analysing Monte Carlo samples}},
  \href{http://arxiv.org/abs/1910.13970}{{\tt arXiv:1910.13970}}.

\bibitem{Chluba:2019kpb}
J.~Chluba et~al., {\it {Spectral Distortions of the CMB as a Probe of
  Inflation, Recombination, Structure Formation and Particle Physics}:
  {Astro2020 Science White Paper}},  {\em Bull. Am. Astron. Soc.} {\bf 51}
  (2019), no.~3 184, [\href{http://arxiv.org/abs/1903.04218}{{\tt
  arXiv:1903.04218}}].

\bibitem{Planck:2018jri}
{\bf Planck} Collaboration, Y.~Akrami et~al., {\it {Planck 2018 results. X.
  Constraints on inflation}},  {\em Astron. Astrophys.} {\bf 641} (2020) A10,
  [\href{http://arxiv.org/abs/1807.06211}{{\tt arXiv:1807.06211}}].

\bibitem{dodelson}
S.~Dodelson, {\em {Modern Cosmology}}.
\newblock Academic Press, Elsevier Science, 2003.

\bibitem{Murgia:2017lwo}
R.~Murgia, A.~Merle, M.~Viel, M.~Totzauer, and A.~Schneider, {\it {''Non-cold''
  dark matter at small scales: a general approach}},  {\em JCAP} {\bf 11}
  (2017) 046, [\href{http://arxiv.org/abs/1704.07838}{{\tt arXiv:1704.07838}}].

\bibitem{bi1997evolution}
H.~Bi and A.~F. Davidsen, {\it Evolution of structure in the intergalactic
  medium and the nature of the ly$\alpha$ forest},  {\em The Astrophysical
  Journal} {\bf 479} (1997), no.~2 523.

\bibitem{hui1997equation}
L.~Hui and N.~Y. Gnedin, {\it Equation of state of the photoionized
  intergalactic medium},  {\em Monthly Notices of the Royal Astronomical
  Society} {\bf 292} (1997), no.~1 27--42.

\bibitem{choudhury2001semianalytic}
T.~R. Choudhury, R.~Srianand, and T.~Padmanabhan, {\it Semianalytic approach to
  understanding the distribution of neutral hydrogen in the universe:
  comparison of simulations with observations},  {\em The Astrophysical
  Journal} {\bf 559} (2001), no.~1 29.

\bibitem{pandey2012probing}
K.~L. Pandey and S.~K. Sethi, {\it Probing primordial magnetic fields using
  ly$\alpha$ clouds},  {\em The Astrophysical Journal} {\bf 762} (2012), no.~1
  15.

\bibitem{sarkar2021using}
A.~K. Sarkar, K.~L. Pandey, and S.~K. Sethi, {\it Using the redshift evolution
  of the lyman-$\alpha$ effective opacity as a probe of dark matter models},
  {\em Journal of Cosmology and Astroparticle Physics} {\bf 2021} (2021),
  no.~10 077.

\bibitem{2011MNRAS.413.1717B}
S.~{Bird}, H.~V. {Peiris}, M.~{Viel}, and L.~{Verde}, {\it {Minimally
  parametric power spectrum reconstruction from the Lyman {\ensuremath{\alpha}}
  forest}},  {\em MNRAS} {\bf 413} (May, 2011) 1717--1728,
  [\href{http://arxiv.org/abs/1010.1519}{{\tt arXiv:1010.1519}}].

\bibitem{Bringmann:2011ut}
T.~Bringmann, P.~Scott, and Y.~Akrami, {\it {Improved constraints on the
  primordial power spectrum at small scales from ultracompact minihalos}},
  {\em Phys. Rev. D} {\bf 85} (2012) 125027,
  [\href{http://arxiv.org/abs/1110.2484}{{\tt arXiv:1110.2484}}].

\bibitem{Kohri:2014lza}
K.~Kohri, T.~Nakama, and T.~Suyama, {\it {Testing scenarios of primordial black
  holes being the seeds of supermassive black holes by ultracompact minihalos
  and CMB $\mu$-distortions}},  {\em Phys. Rev. D} {\bf 90} (2014), no.~8
  083514, [\href{http://arxiv.org/abs/1405.5999}{{\tt arXiv:1405.5999}}].

\bibitem{Nakama:2017qac}
T.~Nakama, T.~Suyama, K.~Kohri, and N.~Hiroshima, {\it {Constraints on
  small-scale primordial power by annihilation signals from extragalactic dark
  matter minihalos}},  {\em Phys. Rev. D} {\bf 97} (2018), no.~2 023539,
  [\href{http://arxiv.org/abs/1712.08820}{{\tt arXiv:1712.08820}}].

\bibitem{2011JCAP...07..025K}
A.~{Kogut}, D.~J. {Fixsen}, D.~T. {Chuss}, J.~{Dotson}, E.~{Dwek},
  M.~{Halpern}, G.~F. {Hinshaw}, S.~M. {Meyer}, S.~H. {Moseley}, M.~D.
  {Seiffert}, D.~N. {Spergel}, and E.~J. {Wollack}, {\it {The Primordial
  Inflation Explorer (PIXIE): a nulling polarimeter for cosmic microwave
  background observations}},  {\em JCAP} {\bf 2011} (July, 2011) 025,
  [\href{http://arxiv.org/abs/1105.2044}{{\tt arXiv:1105.2044}}].

\bibitem{Cai:2018dig}
R.-g. Cai, S.~Pi, and M.~Sasaki, {\it {Gravitational Waves Induced by
  non-Gaussian Scalar Perturbations}},  {\em Phys. Rev. Lett.} {\bf 122}
  (2019), no.~20 201101, [\href{http://arxiv.org/abs/1810.11000}{{\tt
  arXiv:1810.11000}}].

\bibitem{Unal:2018yaa}
C.~Unal, {\it {Imprints of Primordial Non-Gaussianity on Gravitational Wave
  Spectrum}},  {\em Phys. Rev. D} {\bf 99} (2019), no.~4 041301,
  [\href{http://arxiv.org/abs/1811.09151}{{\tt arXiv:1811.09151}}].

\bibitem{Atal:2021jyo}
V.~Atal and G.~Dom\`enech, {\it {Probing non-Gaussianities with the high
  frequency tail of induced gravitational waves}},  {\em JCAP} {\bf 06} (2021)
  001, [\href{http://arxiv.org/abs/2103.01056}{{\tt arXiv:2103.01056}}].
  [Erratum: JCAP 10, E01 (2023)].

\bibitem{Adshead:2021hnm}
P.~Adshead, K.~D. Lozanov, and Z.~J. Weiner, {\it {Non-Gaussianity and the
  induced gravitational wave background}},  {\em JCAP} {\bf 10} (2021) 080,
  [\href{http://arxiv.org/abs/2105.01659}{{\tt arXiv:2105.01659}}].

\bibitem{Zhang:2021vak}
F.~Zhang, J.~Lin, and Y.~Lu, {\it {Double-peaked inflation model: Scalar
  induced gravitational waves and primordial-black-hole suppression from
  primordial non-Gaussianity}},  {\em Phys. Rev. D} {\bf 104} (2021), no.~6
  063515, [\href{http://arxiv.org/abs/2106.10792}{{\tt arXiv:2106.10792}}].
  [Erratum: Phys.Rev.D 104, 129902 (2021)].

\bibitem{Ragavendra:2021qdu}
H.~V. Ragavendra, {\it {Accounting for scalar non-Gaussianity in secondary
  gravitational waves}},  {\em Phys. Rev. D} {\bf 105} (2022), no.~6 063533,
  [\href{http://arxiv.org/abs/2108.04193}{{\tt arXiv:2108.04193}}].

\bibitem{Garcia-Saenz:2022tzu}
S.~Garcia-Saenz, L.~Pinol, S.~Renaux-Petel, and D.~Werth, {\it {No-go theorem
  for scalar-trispectrum-induced gravitational waves}},  {\em JCAP} {\bf 03}
  (2023) 057, [\href{http://arxiv.org/abs/2207.14267}{{\tt arXiv:2207.14267}}].

\bibitem{Chen:2022dah}
C.~Chen, A.~Ota, H.-Y. Zhu, and Y.~Zhu, {\it {Missing one-loop contributions in
  secondary gravitational waves}},  \href{http://arxiv.org/abs/2210.17176}{{\tt
  arXiv:2210.17176}}.

\bibitem{Li:2023qua}
J.-P. Li, S.~Wang, Z.-C. Zhao, and K.~Kohri, {\it {Primordial non-Gaussianity f
  $_{NL}$ and anisotropies in scalar-induced gravitational waves}},  {\em JCAP}
  {\bf 10} (2023) 056, [\href{http://arxiv.org/abs/2305.19950}{{\tt
  arXiv:2305.19950}}].

\bibitem{Franciolini:2023pbf}
G.~Franciolini, A.~Iovino, Junior., V.~Vaskonen, and H.~Veermae, {\it {Recent
  Gravitational Wave Observation by Pulsar Timing Arrays and Primordial Black
  Holes: The Importance of Non-Gaussianities}},  {\em Phys. Rev. Lett.} {\bf
  131} (2023), no.~20 201401, [\href{http://arxiv.org/abs/2306.17149}{{\tt
  arXiv:2306.17149}}].

\bibitem{Liu:2023ymk}
L.~Liu, Z.-C. Chen, and Q.-G. Huang, {\it {Implications for the non-Gaussianity
  of curvature perturbation from pulsar timing arrays}},  {\em Phys. Rev. D}
  {\bf 109} (2024), no.~6 L061301, [\href{http://arxiv.org/abs/2307.01102}{{\tt
  arXiv:2307.01102}}].

\bibitem{Yuan:2023ofl}
C.~Yuan, D.-S. Meng, and Q.-G. Huang, {\it {Full analysis of the scalar-induced
  gravitational waves for the curvature perturbation with local-type
  non-Gaussianities}},  {\em JCAP} {\bf 12} (2023) 036,
  [\href{http://arxiv.org/abs/2308.07155}{{\tt arXiv:2308.07155}}].

\bibitem{Li:2023xtl}
J.-P. Li, S.~Wang, Z.-C. Zhao, and K.~Kohri, {\it {Complete Analysis of
  Scalar-Induced Gravitational Waves and Primordial Non-Gaussianities
  $f_{\mathrm{NL}}$ and $g_{\mathrm{NL}}$}},
  \href{http://arxiv.org/abs/2309.07792}{{\tt arXiv:2309.07792}}.

\bibitem{Choudhury:2023fwk}
S.~Choudhury, K.~Dey, A.~Karde, S.~Panda, and M.~Sami, {\it {Primordial
  non-Gaussianity as a saviour for PBH overproduction in SIGWs generated by
  Pulsar Timing Arrays for Galileon inflation}},
  \href{http://arxiv.org/abs/2310.11034}{{\tt arXiv:2310.11034}}.

\bibitem{Chang:2023aba}
Z.~Chang, Y.-T. Kuang, D.~Wu, J.-Z. Zhou, and Q.-H. Zhu, {\it {New constraints
  on primordial non-Gaussianity from missing two-loop contributions of scalar
  induced gravitational waves}},  {\em Phys. Rev. D} {\bf 109} (2024), no.~4
  L041303, [\href{http://arxiv.org/abs/2311.05102}{{\tt arXiv:2311.05102}}].

\bibitem{Inui:2023qsd}
R.~Inui, S.~Jaraba, S.~Kuroyanagi, and S.~Yokoyama, {\it {Constraints on
  Non-Gaussian primordial curvature perturbation from the LIGO-Virgo-KAGRA
  third observing run}},  \href{http://arxiv.org/abs/2311.05423}{{\tt
  arXiv:2311.05423}}.

\bibitem{Chang:2023vjk}
Z.~Chang, Y.-T. Kuang, D.~Wu, and J.-Z. Zhou, {\it {Probing scalar induced
  gravitational waves with PTA and LISA: The Importance of third order
  correction}},  \href{http://arxiv.org/abs/2312.14409}{{\tt
  arXiv:2312.14409}}.

\bibitem{Papanikolaou:2024kjb}
T.~Papanikolaou, X.-C. He, X.-H. Ma, Y.-F. Cai, E.~N. Saridakis, and M.~Sasaki,
  {\it {New probe of non-Gaussianities with primordial black hole induced
  gravitational waves}},  \href{http://arxiv.org/abs/2403.00660}{{\tt
  arXiv:2403.00660}}.

\bibitem{2023MNRAS.518.6011D}
C.~T. {Donnan}, D.~J. {McLeod}, J.~S. {Dunlop}, R.~J. {McLure}, A.~C.
  {Carnall}, R.~{Begley}, F.~{Cullen}, M.~L. {Hamadouche}, R.~A.~A. {Bowler},
  D.~{Magee}, H.~J. {McCracken}, B.~{Milvang-Jensen}, A.~{Moneti}, and
  T.~{Targett}, {\it {The evolution of the galaxy UV luminosity function at
  redshifts z $\simeq$ 8 - 15 from deep JWST and ground-based near-infrared
  imaging}},  {\em Monthly Notices of the Royal Astronomical Society} {\bf 518}
  (Feb., 2023) 6011--6040, [\href{http://arxiv.org/abs/2207.12356}{{\tt
  arXiv:2207.12356}}].

\bibitem{2023Natur.616..266L}
I.~{Labb{\'e}}, P.~{van Dokkum}, E.~{Nelson}, R.~{Bezanson}, K.~A. {Suess},
  J.~{Leja}, G.~{Brammer}, K.~{Whitaker}, E.~{Mathews}, M.~{Stefanon}, and
  B.~{Wang}, {\it {A population of red candidate massive galaxies 600 Myr after
  the Big Bang}},  {\em Nature} {\bf 616} (Apr., 2023) 266--269,
  [\href{http://arxiv.org/abs/2207.12446}{{\tt arXiv:2207.12446}}].

\bibitem{Syu:2019uwx}
W.-C. Syu, D.-S. Lee, and K.-W. Ng, {\it {Quantum loop effects to the power
  spectrum of primordial perturbations during ultra slow-roll inflation}},
  {\em Phys. Rev. D} {\bf 101} (2020), no.~2 025013,
  [\href{http://arxiv.org/abs/1907.13089}{{\tt arXiv:1907.13089}}].

\bibitem{Kristiano:2022maq}
J.~Kristiano and J.~Yokoyama, {\it {Ruling Out Primordial Black Hole Formation
  From Single-Field Inflation}},  \href{http://arxiv.org/abs/2211.03395}{{\tt
  arXiv:2211.03395}}.

\bibitem{Inomata:2022yte}
K.~Inomata, M.~Braglia, X.~Chen, and S.~Renaux-Petel, {\it {Questions on
  calculation of primordial power spectrum with large spikes: the resonance
  model case}},  {\em JCAP} {\bf 04} (2023) 011,
  [\href{http://arxiv.org/abs/2211.02586}{{\tt arXiv:2211.02586}}]. [Erratum:
  JCAP 09, E01 (2023)].

\bibitem{Choudhury:2023vuj}
S.~Choudhury, M.~R. Gangopadhyay, and M.~Sami, {\it {No-go for the formation of
  heavy mass Primordial Black Holes in Single Field Inflation}},
  \href{http://arxiv.org/abs/2301.10000}{{\tt arXiv:2301.10000}}.

\bibitem{Firouzjahi:2023bkt}
H.~Firouzjahi, {\it {Revisiting Loop Corrections in Single Field USR
  Inflation}},  \href{http://arxiv.org/abs/2311.04080}{{\tt arXiv:2311.04080}}.

\bibitem{Cheng:2023ikq}
S.-L. Cheng, D.-S. Lee, and K.-W. Ng, {\it {Primordial perturbations from
  ultra-slow-roll single-field inflation with quantum loop effects}},
  \href{http://arxiv.org/abs/2305.16810}{{\tt arXiv:2305.16810}}.

\bibitem{Choudhury:2023rks}
S.~Choudhury, S.~Panda, and M.~Sami, {\it {Quantum loop effects on the power
  spectrum and constraints on primordial black holes}},
  \href{http://arxiv.org/abs/2303.06066}{{\tt arXiv:2303.06066}}.

\bibitem{Maity:2023qzw}
S.~Maity, H.~V. Ragavendra, S.~K. Sethi, and L.~Sriramkumar, {\it {Loop
  contributions to the scalar power spectrum due to quartic order action in
  ultra slow roll inflation}},  \href{http://arxiv.org/abs/2307.13636}{{\tt
  arXiv:2307.13636}}.

\bibitem{Riotto:2023hoz}
A.~Riotto, {\it {The Primordial Black Hole Formation from Single-Field
  Inflation is Not Ruled Out}},  \href{http://arxiv.org/abs/2301.00599}{{\tt
  arXiv:2301.00599}}.

\bibitem{Firouzjahi:2023ahg}
H.~Firouzjahi and A.~Riotto, {\it {Primordial Black Holes and Loops in
  Single-Field Inflation}},  \href{http://arxiv.org/abs/2304.07801}{{\tt
  arXiv:2304.07801}}.

\bibitem{Fumagalli:2023hpa}
J.~Fumagalli, {\it {Absence of one-loop effects on large scales from small
  scales in non-slow-roll dynamics}},
  \href{http://arxiv.org/abs/2305.19263}{{\tt arXiv:2305.19263}}.

\bibitem{Cheng:2021lif}
S.-L. Cheng, D.-S. Lee, and K.-W. Ng, {\it {Power spectrum of primordial
  perturbations during ultra-slow-roll inflation with back reaction effects}},
  {\em Phys. Lett. B} {\bf 827} (2022) 136956,
  [\href{http://arxiv.org/abs/2106.09275}{{\tt arXiv:2106.09275}}].

\end{thebibliography}\endgroup

\end{document}